%% file: Knierman_aph.tex
\documentclass[preprint2]{aastex}
\newcommand{\etal}{et~al.\ }
\newcommand{\PVdblt}{{\rm P}\kern 0.1em{\sc v}~$\lambda\lambda 1117, 1128$}
\newcommand{\CaIIdblt}{{\rm Ca}\kern 0.1em{\sc ii}~$\lambda\lambda 3934, 3969$}
\newcommand{\AlIIIdblt}{{\rm Al}\kern 0.1em{\sc iii}~$\lambda\lambda 1854, 1862$}
\newcommand{\CIVdblt}{{\rm C}\kern 0.1em{\sc iv}~$\lambda\lambda 1548, 1550$}
\newcommand{\MgIIdblt}{{\rm Mg}\kern 0.1em{\sc ii}~$\lambda\lambda 2796, 2803$}
\newcommand{\NVdblt}{{\rm N}\kern 0.1em{\sc v}~$\lambda\lambda 1238, 1242$}  
\newcommand{\SVIdblt}{{\rm S}\kern 0.1em{\sc vi}~$\lambda\lambda 933, 944$} 
\newcommand{\OVIdblt}{{\rm O}\kern 0.1em{\sc vi}~$\lambda\lambda 1031, 1037$} 
\newcommand{\SiIIdblt}{{\rm Si}\kern 0.1em{\sc ii}~$\lambda\lambda 1190, 1193$} 
\newcommand{\SiIVdblt}{{\rm Si}\kern 0.1em{\sc iv}~$\lambda\lambda 1393, 1402$} 
\newcommand{\PV}{\hbox{{\rm P}\kern 0.1em{\sc v}}}
\newcommand{\AlI}{\hbox{{\rm Al}\kern 0.1em{\sc i}}}
\newcommand{\AlII}{\hbox{{\rm Al}\kern 0.1em{\sc ii}}}
\newcommand{\AlIII}{{\hbox{\rm Al}\kern 0.1em{\sc iii}}}
\newcommand{\CaII}{\hbox{{\rm Ca}\kern 0.1em{\sc ii}}}
\newcommand{\CII}{\hbox{{\rm C}\kern 0.1em{\sc ii}}}
\newcommand{\CIIe}{\hbox{{\rm C$^{\ast}$}\kern 0.1em{\sc ii}}}
\newcommand{\CIII}{\hbox{{\rm C}\kern 0.1em{\sc iii}}}
\newcommand{\CIV}{\hbox{{\rm C}\kern 0.1em{\sc iv}}}
\newcommand{\CV}{\hbox{{\rm C}\kern 0.1em{\sc v}}}
\newcommand{\HI}{\hbox{{\rm H}\kern 0.1em{\sc i}}}
\newcommand{\HII}{\hbox{{\rm H}\kern 0.1em{\sc ii}}}
\newcommand{\Lya}{\hbox{{\rm Ly}\kern 0.1em$\alpha$}}
\newcommand{\Ha}{\hbox{{\rm H}\kern 0.1em$\alpha$}}
\newcommand{\Lyb}{\hbox{{\rm Ly}\kern 0.1em$\beta$}}
\newcommand{\Lyg}{\hbox{{\rm Ly}\kern 0.1em$\gamma$}}
\newcommand{\Lyd}{\hbox{{\rm Ly}\kern 0.1em$\delta$}}
\newcommand{\Lye}{\hbox{{\rm Ly}\kern 0.1em$\epsilon$}}
\newcommand{\Lyphi}{\hbox{{\rm Ly}\kern 0.1em$\phi$}}
\newcommand{\Lyfive}{\hbox{{\rm Ly}\kern 0.1em$5$}}
\newcommand{\Lysix}{\hbox{{\rm Ly}\kern 0.1em$6$}}
\newcommand{\Lyseven}{\hbox{{\rm Ly}\kern 0.1em$7$}}
\newcommand{\Lyeight}{\hbox{{\rm Ly}\kern 0.1em$8$}}
\newcommand{\Lynine}{\hbox{{\rm Ly}\kern 0.1em$9$}}
\newcommand{\Lyten}{\hbox{{\rm Ly}\kern 0.1em$10$}}
\newcommand{\Lyeleven}{\hbox{{\rm Ly}\kern 0.1em$11$}}
\newcommand{\HeI}{\hbox{{\rm He}\kern 0.1em{\sc i}}}
\newcommand{\HeII}{\hbox{{\rm He}\kern 0.1em{\sc ii}}}
\newcommand{\FeI}{\hbox{{\rm Fe}\kern 0.1em{\sc i}}}
\newcommand{\FeII}{\hbox{{\rm Fe}\kern 0.1em{\sc ii}}}
\newcommand{\FeIII}{\hbox{{\rm Fe}\kern 0.1em{\sc iii}}}
\newcommand{\MnII}{\hbox{{\rm Mn}\kern 0.1em{\sc ii}}}
\newcommand{\MgI}{\hbox{{\rm Mg}\kern 0.1em{\sc i}}}
\newcommand{\MgII}{\hbox{{\rm Mg}\kern 0.1em{\sc ii}}}
\newcommand{\MgIII}{\hbox{{\rm Mg}\kern 0.1em{\sc iii}}}
\newcommand{\NI}{\hbox{{\rm N}\kern 0.1em{\sc i}}}
\newcommand{\NII}{\hbox{{\rm N}\kern 0.1em{\sc ii}}}
\newcommand{\NIII}{\hbox{{\rm N}\kern 0.1em{\sc iii}}}
\newcommand{\NV}{\hbox{{\rm N}\kern 0.1em{\sc v}}}
\newcommand{\OVI}{\hbox{{\rm O}\kern 0.1em{\sc vi}}}
\newcommand{\OI}{\hbox{{\rm O}\kern 0.1em{\sc i}}}
\newcommand{\OII}{\hbox{[{\rm O}\kern 0.1em{\sc ii}]}}
\newcommand{\OIV}{\hbox{{\rm O}\kern 0.1em{\sc iv}]}}
\newcommand{\SI}{{\rm S}\kern 0.1em{\sc i}}
\newcommand{\SIV}{{\rm S}\kern 0.1em{\sc iv}}
\newcommand{\SVI}{{\rm S}\kern 0.1em{\sc vi}}
\newcommand{\SiI}{\hbox{{\rm Si}\kern 0.1em{\sc i}}}
\newcommand{\SiII}{\hbox{{\rm Si}\kern 0.1em{\sc ii}}}
\newcommand{\SiIII}{\hbox{{\rm Si}\kern 0.1em{\sc iii}}}
\newcommand{\SiIV}{\hbox{{\rm Si}\kern 0.1em{\sc iv}}}
\newcommand{\SII}{\hbox{{\rm S}\kern 0.1em{\sc ii}}}
\newcommand{\SIII}{\hbox{{\rm S}\kern 0.1em{\sc iii}}}
\newcommand{\NaI}{\hbox{{\rm Na}\kern 0.1em{\sc i}}}
\newcommand{\TiII}{\hbox{{\rm Ti}\kern 0.1em{\sc ii}}}
\newcommand{\kms}{\hbox{km~s$^{-1}$}}

\newcommand{\msun}{${\rm M}_{\odot}$}
 
\begin{document}
 
%\received{date month year}
%\accepted{date month year}
%\journalid{number}{date month year}
%\articleid{number}{number}
%\slugcomment{The Astronomical Journal, {\rm submitted}}

%%%%%%%%%%%%%%%%%%%%%%%%%%%%%%%%%%%%%%%%%%%%%%%%%%%%%%%%%%%%%%%%%%%%%%%%%%%%%%%%%%
%% - Title and Author List 
%%\title{\altaffilmark{1,2}}

\title{From Globular Clusters to Tidal Dwarfs:  Structure Formation in
the Tidal Tails of Merging Galaxies\altaffilmark{1}}

\author{Karen~A.~Knierman\altaffilmark{2}, Sarah~C.~Gallagher\altaffilmark{3}, Jane~C.~Charlton,
Sally~D.~Hunsberger}
\affil{Department of Astronomy and Astrophysics \\ 
       The Pennsylvania State University \\
       University Park PA 16802 \\ 
      {\it kak, gallsc, charlton, sdh@astro.psu.edu}}

\author{Bradley Whitmore}
\affil{Space Telescope Science Institute \\
       3700 San Martin Drive \\
        Baltimore, MD 21218 \\
      {\it whitmore@stsci.edu}}

\author{Arunav Kundu}
\affil{Physics and Astronomy Dept \\
       Michigan State University \\
       East Lansing, MI 48824 \\
      {\it akundu@pa.msu.edu}}

\author{J.~E. Hibbard}
\affil{National Radio Astronomy Observatory \\
       520 Edgemont Road \\
       Charlottesville, VA 22903 \\
      {\it jhibbard@nrao.edu}}
\and
\author{Dennis Zaritsky}
\affil{Steward Observatory \\
       University of Arizona \\
       Tucson, AZ 85721 \\
      {\it dennis@ngala.as.arizona.edu}}

\altaffiltext{1}{Based in part on observations obtained with the
NASA/ESA {\it Hubble Space Telescope}, which is operated by the STScI
for the Association of Universities for Research in Astronomy, Inc.,
under NASA contract NAS5--26555.}
\altaffiltext{2}{Now at Steward Obs., Univ. of Arizona, Tucson, AZ 85721;
kknierman@as.arizona.edu}
\altaffiltext{3}{Now at Center for Space Research, Massachusetts
Institute of Technology, 77 Massachusetts Avenue, Cambridge, MA 02139; 
scg@space.mit.edu}

\begin{abstract}

Using $V$ and $I$ images obtained with the Wide Field Planetary Camera
2 (WFPC2) of the {\it Hubble Space Telescope}, we investigate compact
stellar structures within tidal tails.  Six regions of tidal debris in
the four classic ``Toomre Sequence'' mergers: NGC 4038/39
(``Antennae''), NGC 3256, NGC 3921, and NGC 7252 (``Atoms for Peace'')
have been studied in order to explore how the star formation depends
upon the local and global physical conditions.  These mergers sample a
range of stages in the evolutionary sequence and tails with and
without embedded tidal dwarf galaxies.  The six tails are found to
contain a variety of stellar structures, with sizes ranging from those
of globular clusters up to those of dwarf galaxies.  From $V$ and $I$
WFPC2 images, we measure the luminosities and colors of the star
clusters.  NGC 3256 is found to have a large population of blue
clusters ($0.2 \la V-I \la 0.9$), particularly in its Western tail,
similar to those found in the inner region of the merger.  In
contrast, NGC 4038/39 has no clusters in the observed region of the
tail, only less luminous point sources likely to be individual stars.
NGC 3921 and NGC 7252 have small populations of clusters along their
tails.  A significant cluster population is clearly associated with
the prominent tidal dwarf candidates in the eastern and western tails
of NGC 7252.  The cluster--rich Western tail of NGC 3256 is {\it not}
distinguished from the others by its dynamical age or by its total
{\HI} mass.  However, the mergers that have few clusters in the tail
all have tidal dwarf galaxies, while NGC 3256 does not have prominent
tidal dwarfs.  We speculate that star formation in tidal tails may
manifest itself either in small structures like clusters along the
tail or in large structures such as dwarf galaxies, but not in
both. Also, NGC 3256 has the highest star formation rate of the four
mergers studied, which may contribute to the high number of star
clusters in its tidal tails.

\end{abstract}

\keywords{galaxies:individual(NGC 4038/39, NGC 3256, NGC 3921, NGC
7252) --- galaxies:interactions --- galaxies:star clusters}

%%%%%%%%%%%%%%%%%%%%%%%%%%%%%%%%%%%%%%%%%%%%%%%%%%%%%%%%%%%%%%%%%%%%%%%%%%%%%%%%%%
%% - section: Introduction
%%   figures: none
%%   tables : none

%\newpage

\section{Introduction}
\label{sec:intro}

There is growing evidence that massive young compact star clusters
form in many different environments. Old globular clusters, like those
seen in the Milky Way galaxy, presumably form in the early stages of
galaxy formation. Recently, young compact clusters were discovered to
be forming in several environments: starburst galaxies \citep{Meurer},
barred galaxies \citep{barth95}, some spiral disks \citep{larsen99},
the inner regions of mergers \citep{Holtz92, Whit93, setal96,
miller97, whit99, zepf99} and in the space between galaxies in compact
groups \citep{sarah}.  Finding common characteristics between these
environments may allow us to identify the mechanism that causes gas
clouds to collapse and efficient star formation to begin
\citep[e.g.,][]{Jog92}. Also, there may be differences from
environment to environment that lead to different types of
``packaging'' of new stars, from isolated stars, to small groups, to
globular clusters, to stellar associations, to dwarf galaxies.  To
understand the underlying conditions responsible for the formation of
these structures, it is important to identify the different
environments in which they form.

Star clusters tend to form wherever vigorous star formation occurs
and, especially, in starbursts triggered by galaxy interactions and
mergers \citep{S98}.  Studies with {\it Hubble Space Telescope} (HST)
of several mergers and merger remnants show a large number of young
clusters in the inner regions of the galaxies \citep{Whit93, Whit95,
setal96, miller97, whit99, zepf99}.  The clusters have effective radii
comparable to, but perhaps slightly larger than, the old globular
clusters in the Milky Way.  Their luminosity functions are best
described as power laws ($\alpha \sim -2$) when expressed as the
number of clusters per luminosity bin.  While this is similar to the
luminosity function of globular clusters in the Milky Way for masses
greater than $\sim 10^5$~\msun \citep{harpud}, merging galaxies have
many more low mass clusters than found in galaxies with old globular
cluster populations.  This disparity could indicate a difference in
the cluster formation process in primordial galaxies as compared to
modern merger remnants, or it could simply indicate that evolution of
the luminosity function has occurred due to fading and selective
disruption \citep{zhang99}.

Bridges and tails may also form during strong tidal interactions
involving disk galaxies \citep{Toomre72, Barnes88}.  These tidal
features have blue $UBVR$ colors \citep[e.g., $B-V = 0.53 \pm 0.13$
from Schombert, Wallin, \& Struck--Marcell 1990\nocite{schombert90};
see also][]{weilbacher}, and regions of {\OII} and {\Ha} line emission
have been found within the tails \citep{Schw78, Mirabel92, Duc1, Duc2,
Iglesias01, Weilbacher03}.  The inferred ages of the young stars and
giant {\HII} regions are much less than the dynamical age of the
tails, indicating that star formation is occurring within the tails.

Several factors may influence star formation within tidal tails.
First, the stage of the merger may be important.  Secondly, 21--cm
neutral hydrogen ({\HI}) observations indicate that tails are gas rich
\citep{yun94, hib01b}, and the evolution of the {\HI} may play a role
in the star formation history.  Many tidal tails have associated {\HI}
which is falling back toward the center into the remnant.  This infall
of gas may enable star formation to continue for time scales on the
order of $1$~Gyr after tail formation \citep{HM95}.  Finally, tidal
dwarf galaxies are found in tidal tails in a variety of environments,
particularly at the ends of the tails \citep{Mirabel92, duc94,
huns96}.  These same tails that form tidal dwarfs may or may not form
smaller stellar structures such as star clusters.  Recent results from
the early release observations of the Advanced Camera for Surveys
(ACS) on HST have shown young star clusters in the tidal tails of UGC
10214 (``The Tadpole'') \citep{Tran03,deGrijs03} and NGC 4676 (``The
Mice'') \citep{deGrijs03}.

This paper focuses on the occurrence of star clusters in the tidal
tails of merging galaxies.  New $V$ and $I$--band observations of the
tidal tails of four merging pairs obtained with HST/WFPC2 are
presented.  These are compared with previous observations of the inner
regions of the same galaxies, and to each other, in order to consider
the factors that influence the nature of the stellar systems that form
in the tails.  The four mergers chosen for this study span a range of
dynamical states and tails with and without embedded tidal dwarf
galaxies.

The four systems in this study were chosen from the ``Toomre
Sequence'' \citep{Toomre77} of ongoing mergers.  This sequence is an
optically selected ensemble of strongly interacting galaxies
representing a suggested evolutionary sequence of disk--disk mergers,
based on their stellar tidal tail lengths and the separation of the
two nuclei \citep{Toomre77}.  The four systems chosen for this study
span the entire range of the sequence, from strongly disturbed, but
separate, disks likely to merge (NGC 4038/39, ``The Antennae''), to
nearly fully merged disks (NGC 3256), to fully merged systems with
single nuclei and relaxed stellar profiles (NGC 3921 and NGC 7252,
``Atoms for Peace'').

In the following, Section \ref{sec:obsred} describes the HST
observations, their reduction, and object selection and
photometry. Section \ref{sec:pairs} presents the four mergers, with
separate subsections describing the observations, the inner regions,
the results, and a discussion for each.  Section \ref{sec:conc} then
compares the results for the different mergers and summarizes the main
conclusions.

%%%%%%%%%%%%%%%%%%%%%%%%%%%%%%%%%%%%%%%%%%%%%%%%%%%%%%%%%%%%%%%%%%%%%%%%%%%%%%%%%%
\section{Observations and Reductions}
\label{sec:obsred}

\subsection{Observations}
\label{sec:obs}

Observations of six tidal tails were obtained with HST/WFPC2 during
the period Nov. 1998 to Oct. 1999, as listed in the journal of
observations in Table~\ref{tab:obs}.  Total integration times ranged
from $1000$--$2000$~s in $V$ (F555W filter) and in $I$ (F814W filter).
The integrations were chosen to obtain images to a depth similar to
the observations of the inner regions.  A gain of $7 e^-/{\rm DN}$
was used in all cases.  At least two separate exposures in each filter
were obtained to facilitate removal of cosmic rays.  The positions of
the WFPC2 field of view is superimposed on optical images of the four
mergers in Figure~\ref{fig:overview}.  The six $V$--band images are
presented in Figures~\ref{fig:V4038} -- \ref{fig:EV7252}.

Pairs of exposure were taken of all systems except NGC 3256. The
exposures were offset by $0.25${\arcsec} (i.e. $2.5$ pixels on the 
Wide Field Camera [hereafter WF] and $5.5$ pixels on the Planetary 
Camera [hereafter PC]). Pipeline--reduced images for each filter were 
averaged and cleaned using the IRAF\footnote{IRAF is distributed by 
the National Optical Astronomy Observatories, which are operated by 
AURA, Inc., under contract to the NSF.} tasks GCOMBINE. The tasks 
COSMICRAYS was used to remove hot pixels.

\subsection{Object Detection}
\label{sec:detect}

Objects in the images were detected using the DAOFIND task in the
DAOPHOT package using a threshold of $2.0$ counts per object, finding
objects greater than $2\sigma$ above the local sky background.  The
noise was determined from the sky annulus of $5$ to $8$ pixels around
each object identified by DAOFIND and only objects with a
signal--to--noise per object greater than $3$ were retained.  For 
images with multiple exposures, the objects were detected separately 
in each pointing.  Then the two source lists were matched using the 
offsets calculated from bright stars.  Only those sources 
with detections in both the $V$ and $I$ filters were retained.

Only point--like sources are of interest in this paper, as extended
objects will be addressed in a future study.  In addition to possible
star clusters, this method detects foreground stars and compact
background galaxies.  We use a classification based upon the $V-I$
color and a structural parameter.  To obtain a sample of possible star
clusters, we keep only those sources with $V-I < 2.0$.  Old,
metal-poor globular clusters in the Milky Way have $V-I\sim1$
\citep{Reed88}, so this cut would not exclude old clusters.  The most
metal--rich clusters, from observations of the Milky Way and nearby
elliptical galaxies, have $V-I < 1.5$ \citep{peterson93, ajhar94,
Whitet95, kundu01a, kundu01b}.  Stellar population models by
\citet{Lee02} which take into account the effect of the horizontal
branch on colors show that even solar metallicity globular clusters
only have $V-I \sim~1.2--1.3$.  Also, only those sources with $V_{err}
< 0.25$ were retained.  

As in \citet{Whit93} and \citet{miller97}, we calculated the
concentration index, the difference in $V$ magnitude between an
aperture of $0.5$ pixel radius and an aperture of $3.0$ pixel radius,
denoted by $\Delta_V(0.5-3)$.  A sky annulus of $6$ to $9$ pixels was
used to subtract the background.  For each system, we used a different
concentration index.  In the NGC 3256 images, we classified objects
with $\Delta_V(0.5-3) > 2.4$ as diffuse and removed them from the
sample.  The cutoff for the concentration index was raised to $3.0$
for NGC 4038/39, the closest system, to avoid eliminating any possible
clusters.

For NGC 3921 and NGC 7252, we modified the above photometry method to
probe deeper apparent magnitudes for comparison with the central
regions and larger sized objects, such as the stellar associations
found close to the central regions of NGC 3921 \citep{setal96}.  For
these images we used DAOFIND with the threshold set at $1.5$ rather
than $2.0$ as above.  The signal-to-noise was calculated again, this
time using a cutoff of $1.5$ per object. As above, we retained sources
with $V-I < 2.0$ and $V_{err} < 0.25$.  There was no cutoff for
concentration index.  By removing the size criteria for objects in
these mergers, we retained more diffuse sources which we could compare
with the stellar associations from \citet{setal96}.

For each tail, a region in the $V$--band image was manually identified
as "in-tail", corresponding to contiguous regions with $\sim~1$~count
(DN) above the background (corresponding to a $V$ magnitude surface
brightness of $24.3 - 25.7$ mag arcsec$^{-2}$ according to the
specific tail). All other regions were designated as
"out-of-tail''. In Figures 2--7, "in-tail" point sources are indicated
by open circles, while "out-of-tail" point sources are indicated by
open squares.

\subsection{Completeness}
\label{sec:completeness}

Figure~\ref{fig:completelim} shows the completeness fraction on the
wide--field chips for the observations of NGC 3256W.  The completeness
was determined by adding $10,000$ stars ($100$ at a time to avoid
crowding) using the ADDSTAR program within the DAOPHOT package.  The
surface brightness of the tail is so low ($\sim~1$~DN above the
background) that a single background level is sufficient for
characterizing the completeness fractions, rather than the multiple
curves used in other studies with a wide variety of backgrounds
\citep[e.g.,]{whit99}.  We find that the completeness fraction
reaches $\sim 50$\% at $V=25.8$, or $25.4$ corrected for foreground
extinction, consistent with our visual estimate of the completeness
correction from the color magnitude diagram for NGC 3256W (see
Figure~\ref{fig:cmd3256}).  Comparison of the color magnitude diagrams
indicates that the completeness limits for NGC 3921 and NGC 4038/39 are
similar to that for NGC 3256.  For NGC 7252, our slightly relaxed
detection criterion (see \S~\ref{sec:detect}) yields a slightly deeper
completeness limit, by $\sim 0.2$ magnitudes.

For several reasons, few sources were detected on the PC chip in any
of the fields ($8$ in the NGC 3256W PC, $2$ in the NGC 3256E PC, and
none in any others).  First, the fields were positioned for maximal
total coverage of the tail regions, from which it follows that the
tails do not pass through the PC chip.  Second, the area covered by
the PC chip is smaller than a single WF chip.  Finally, in the PC chip
the light is spread over a larger number of pixels, but the read noise
is the same per pixel.  We are working in a regime where read noise is
important, thus the detection limit is not as faint for the PC as for
the WF (see \citet{whit99} for a similar comparison). Therefore, only
WF observations will be discussed in this study.  PC sources have not
been plotted in Figures 2--7.

\subsection{Photometry}
\label{sec:phot}

Photometric zeropoints were adopted from Table 28.1 of Version 3 of
the {\it HST Data Handbook} \citep{Voit97}.  The \citet{Holtz95}
photometric transformation corrections were applied (to convert from
the HST filter system to Johnson--Cousins).

Aperture photometry was performed on all pointlike objects in the
cleaned images using the PHOT task in the APPHOT package.  The radii
of the object aperture, the inner boundary of the background annulus,
and the outer boundary were 2/5/8 pixels for the WF.  For all fields,
the aperture corrections $V=0.28$ and $I=0.31$ were adopted.  These
were an average of those derived for stars in the three WF chips for
NGC 1700 and NGC 3610 \citep{whit97}, which were calibrated using
point sources in the NGC 5139 ($\omega Cen$) and NGC 6752 fields.
Ideally, we would have used cluster candidates in our own fields to
determine aperture corrections, however this procedure was subject to
large uncertainties due to a lack of sufficiently bright sources.  For
the best case, WF2 in the NGC 3256W image, nine bright stars were used
to determine corrections of $0.276\pm0.014$ in $V$ and $0.293\pm0.006$
in $I$.  These are consistent with our adopted values.  For the
nearest pair, using stellar aperture corrections (instead of clusters)
introduce a systematic error of only a couple hundredths of a
magnitude, insignificant for the purpose of this study.

The foreground extinction due to the Galaxy was corrected for each
pair using the $A_B$ values from \citet{Schlegel98} and the reddening
curve from \citet{mathis90}.  The $A_B$ values for each merger are
listed in Table~\ref{tab:prop}.  We corrected for non--optimal
charge--transfer efficiency (CTE) using the formulae of \citet{whc99}.

%%%%%%%%%%%%%%%%%%%%%%%%%%%%%%%%%%%%%%%%%%%%%%%%%%%%%%%%%%%%%%%%%%%%%%%%%%%%%%%%%%

\section{The Four Merging Pairs}
\label{sec:pairs}

For each of the mergers we will summarize past observations of the
inner regions, present new HST observations, and discuss our results.
Figure~\ref{fig:overview} shows optical images of the four mergers,
with {\HI} contours, and with the locations of the six observed WFPC2
fields superimposed.  Table~\ref{tab:prop} presents a comparison of
tail properties including velocity of merger, distance modulus,
extinction in $B$ magnitudes($A_B$), length from the central region
along the tail to the position where the image begins ($l$), physical
scale of tail region encompassed in the image ($\Delta l$), absolute
magnitude for $50$\% completeness limit ($M_{V,50\%}$), {\HI} mass of
tail ($M_{HI,tail}$), and the approximate age of tail (the projected
length of the tail divided by the rotational or escape velocity).  The
mergers are discussed in increasing order of their place in the Toomre
Sequence.

\subsection{NGC 4038/39 (``The Antennae'')}
\label{sec:s4038}

\begin{figure}
\figurenum{1}
\plotone{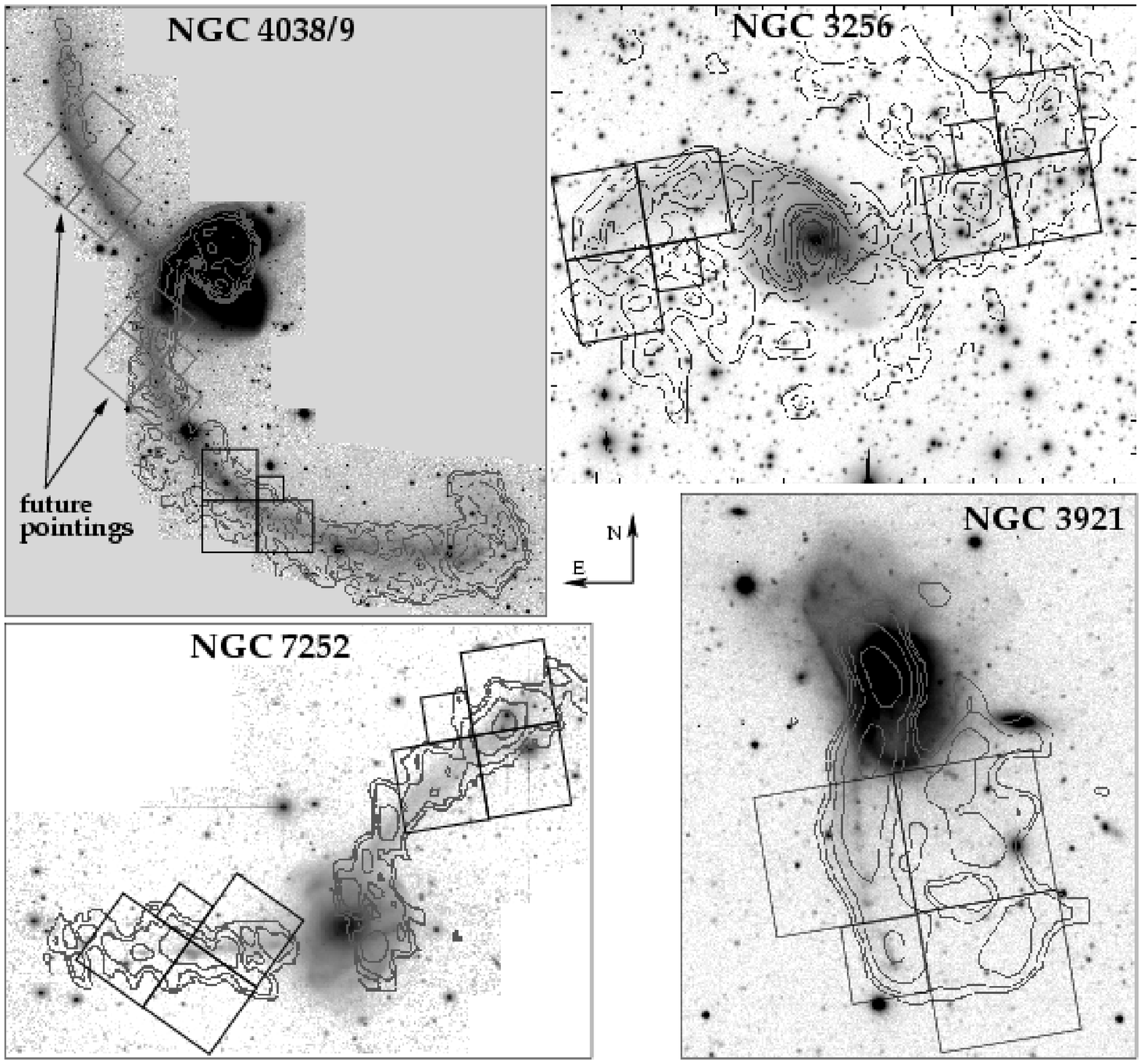}
\caption{Upper left --- Optical image in $BVR$ of early stage remnant NGC
4038/39 (``Antennae'') from the CTIO 0.9m with integrated {\HI} line
emission, shown as overlaid contours taken with the VLA C+D arrays
\citep{hib99}.  The HST WFPC2 field of view of our 1999 Feb 22 observations is
indicated by black boxes, and our two later observations (not analyzed here)
are indicated by gray boxes;  
Upper right --- Optical image of NGC 3256, from UH88'' $R$--band.
The HST WFPC2 fields of view of our observations are indicated by the black boxes.
{\HI} data have been obtained at ATCA \citep{English}.;
Lower left --- Optical image of NGC 7252, from CTIO 4m $R_j$--band, with 
{\HI} contours overlaid \citep{HvG96}.
Lower right --- Optical image of NGC 3921, KPNO 0.9m $R$--band image,
with integrated {\HI} line emission, shown in overlaid contours taken
with the VLA C+D arrays \citep{HvG96}.
\label{fig:overview}}
\end{figure}

\begin{figure}
\figurenum{2}
\plotone{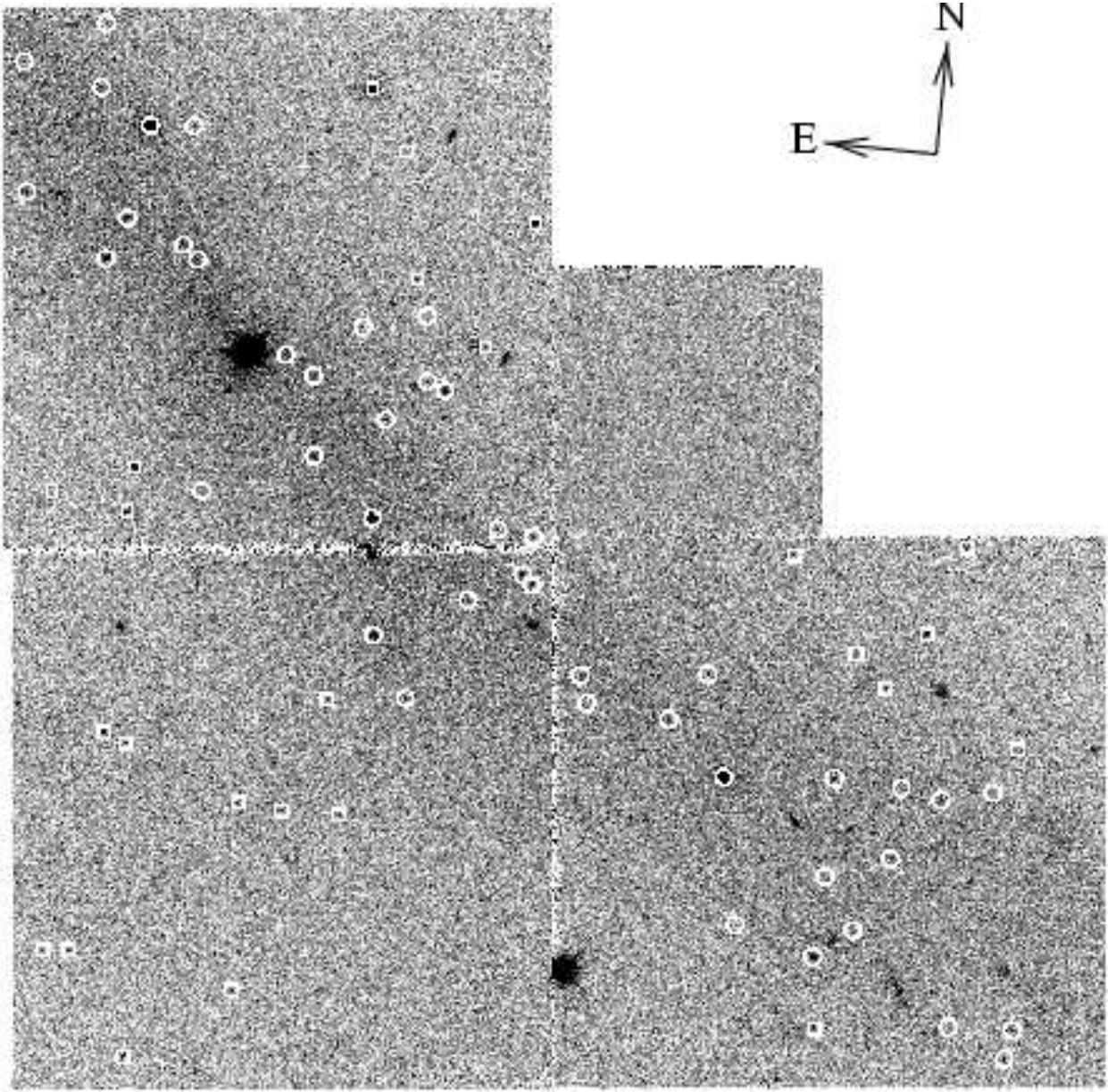}
\caption{
 HST WFPC2 image of NGC 4038/39 (``Antennae'') Southern tail taken with
 F555W ($V$) filter.  Circles indicate sources in the tail and
 squares indicate those outside of the tail.  Sources
 within the tail are likely to be individual stars and not clusters.
\label{fig:V4038}}
\end{figure}

\begin{figure}
\figurenum{3}
\plotone{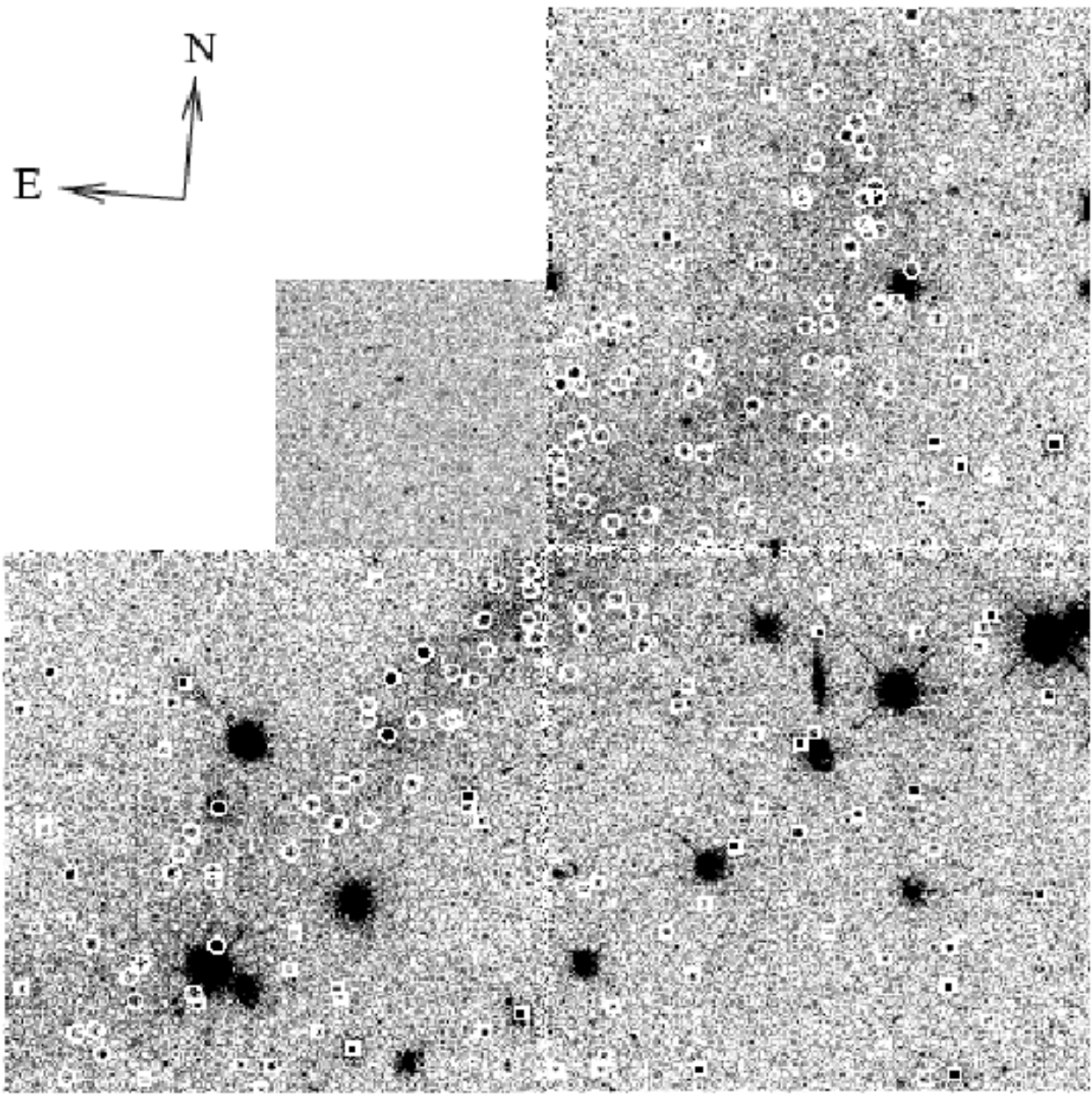}
\caption{
HST WFPC2 image of NGC 3256 Western Tail taken with the F555W ($V$)
filter.  Circles indicate sources within the tail and squares indicate 
those outside the tail.  The majority of sources within the tail are
likely to be clusters; note the large number.
\label{fig:WV3256}}
\end{figure}

\begin{figure}
\figurenum{4}
\plotone{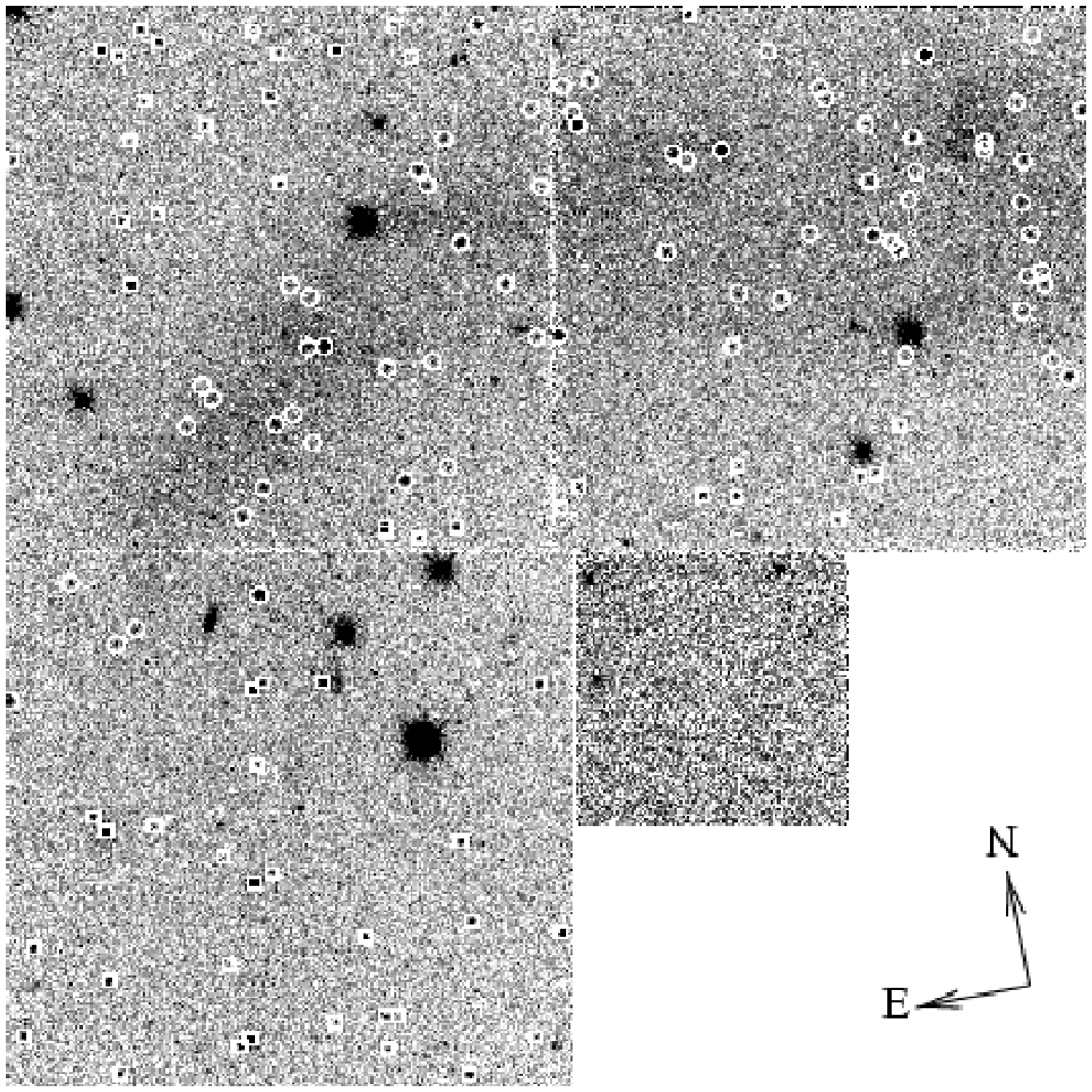}
\caption{
HST WFPC2 image, as in Figure~\ref{fig:WV3256}, for NGC 3256 Eastern
Tail, taken with F555W ($V$) filter.
\label{fig:EV3256}}
\end{figure}

\begin{figure}
\figurenum{5}
\plotone{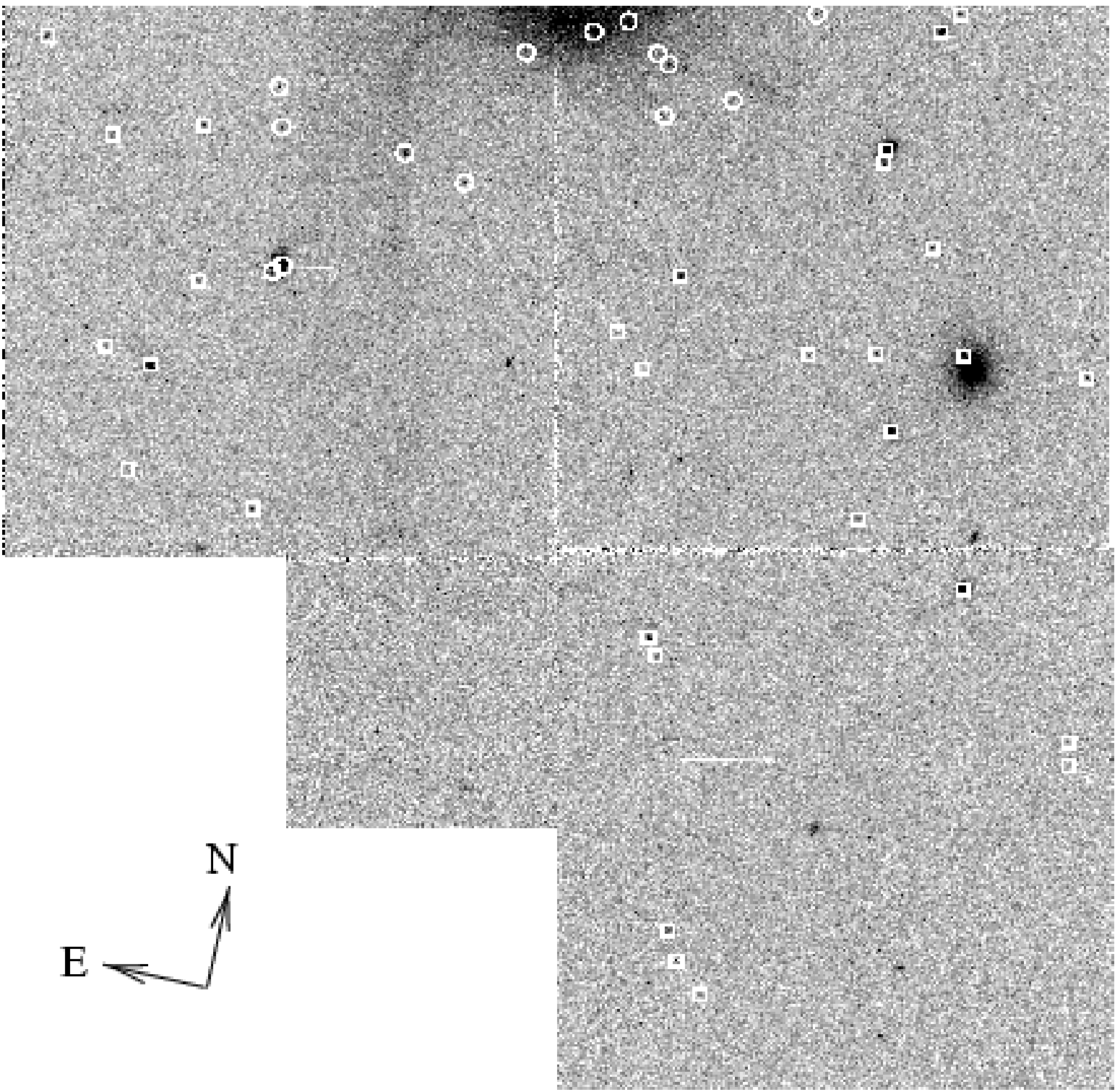}
\caption{
HST WFPC2 image of NGC 3921 Southern Tail, taken with F555W ($V$)
filter. Circles indicate sources within the tail and squares indicate
those outside the tail.
\label{fig:V3921}}
\end{figure}

\begin{figure}
\figurenum{6}
\plotone{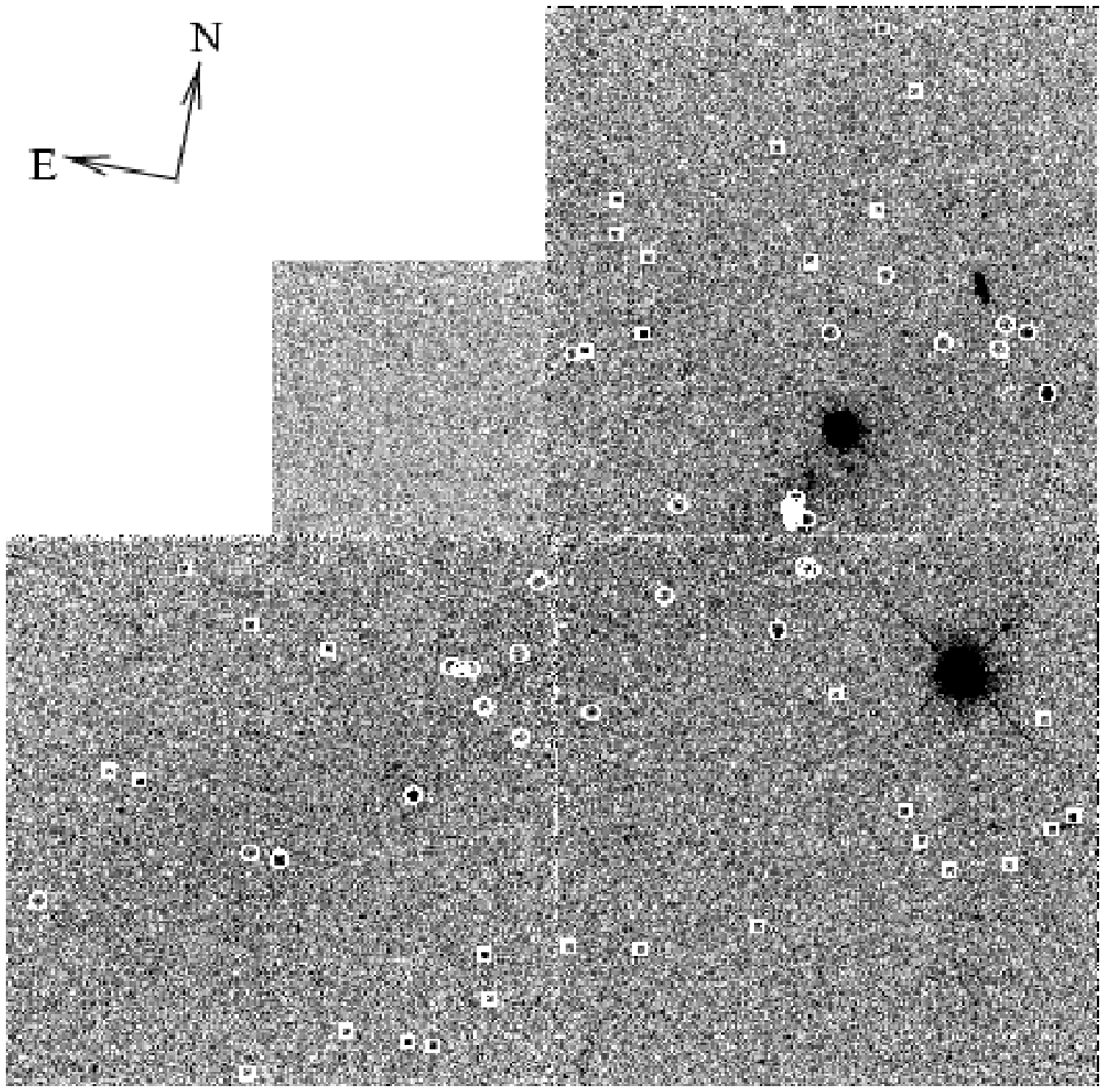}
\caption{
HST WFPC2 image of NGC 7252 Western Tail, taken with F555W ($V$)
filter.  Circles indicate sources within the tail and squares indicate
those outside the tail.
\label{fig:WV7252}}
\end{figure}

\begin{figure}
\figurenum{7}
\plotone{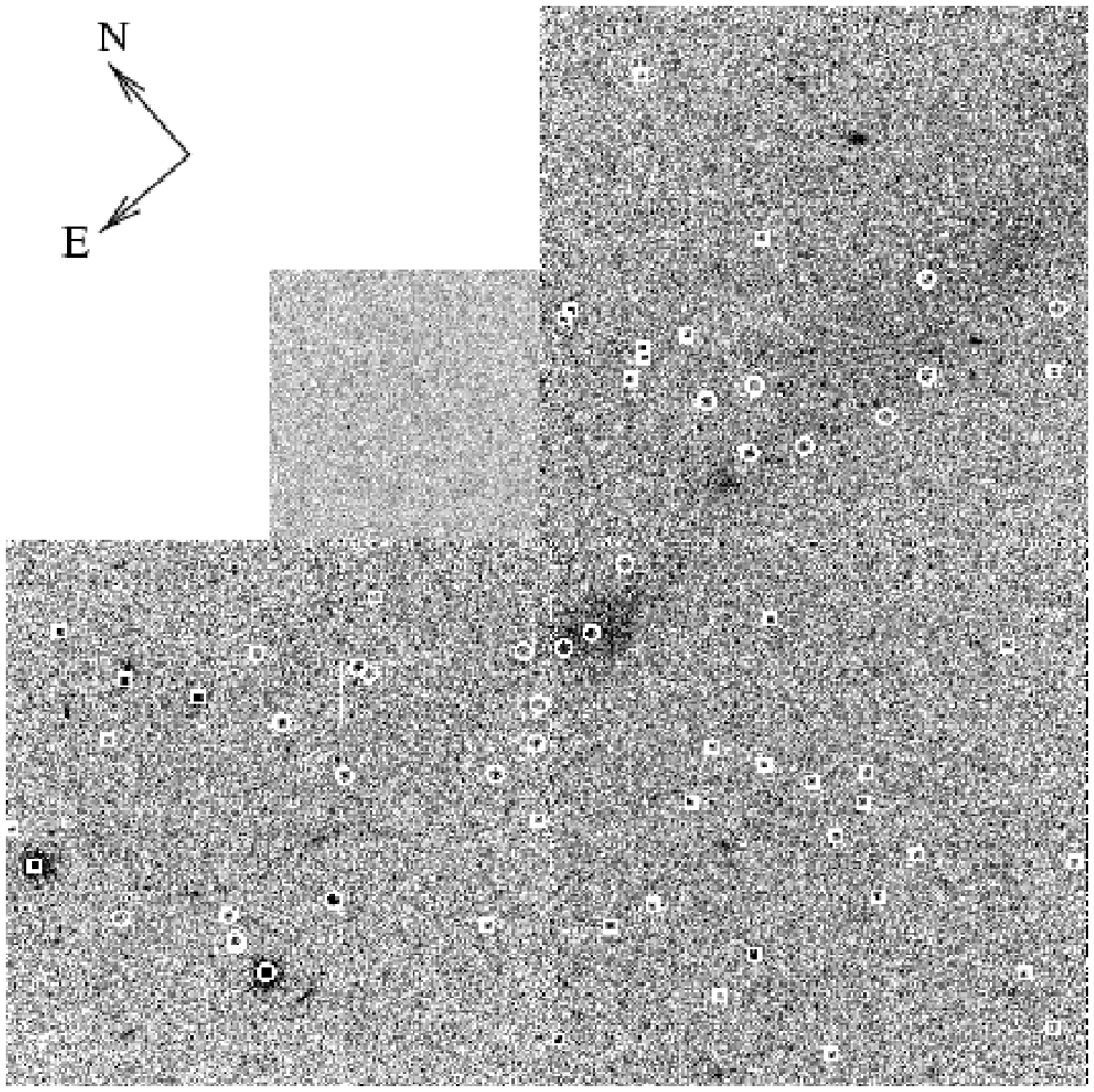}
\caption{
HST WFPC2 image, as in Figure~\ref{fig:WV7252}, for NGC 7252 Eastern
Tail, taken with F555W ($V$) filter.
\label{fig:EV7252}}
\end{figure}

\begin{figure}
\figurenum{8}
\plotone{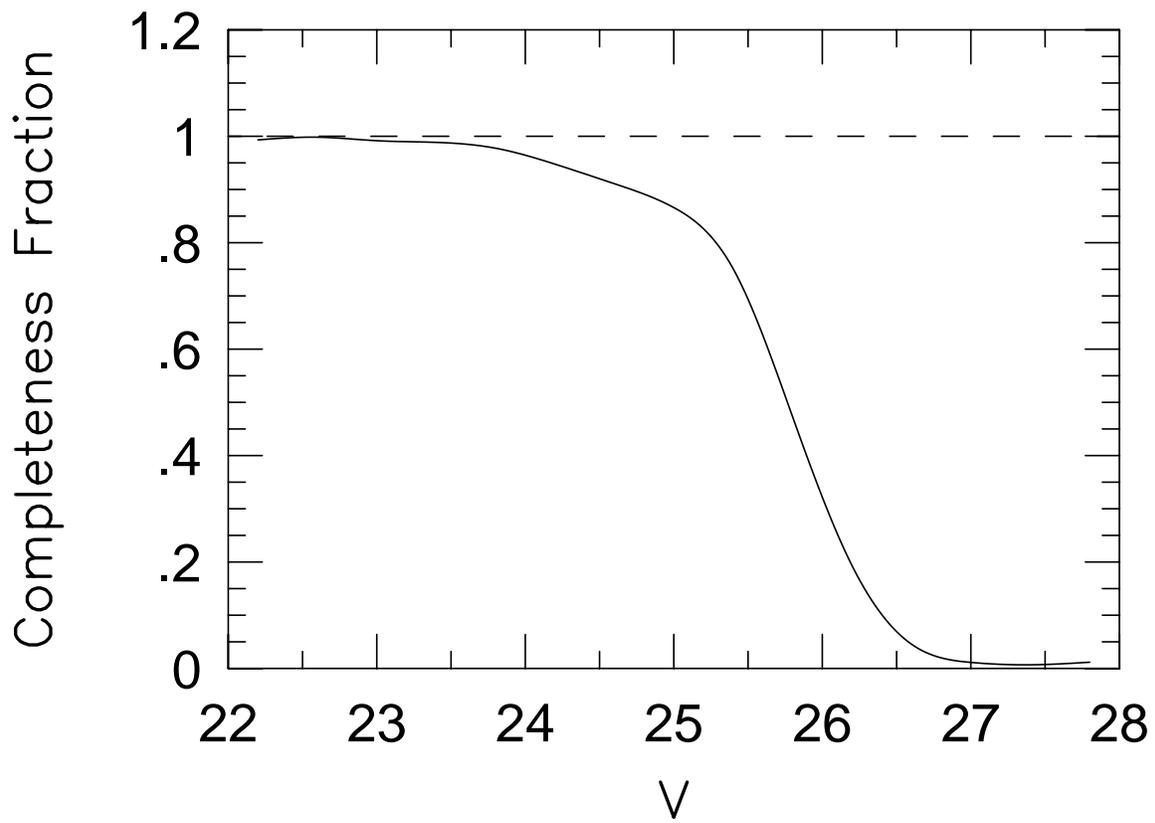}
\caption{
The completeness fraction for point sources as a function of the V magnitude,
determined by adding artificial sources to the WF images of the NGC 3256
Western tail.
\label{fig:completelim}}
\end{figure}

The Antennae is one of the best known and earliest modeled merger
systems \citep{Toomre72, vdH79}.  It is the nearest of the four
mergers in this study (see Table~\ref{tab:prop}).  Its long, nearly
symmetric crossed tidal tails and still--separate nuclei indicate a
relatively young, strongly interacting merger; see
Figure~\ref{fig:overview} for an optical image.  The southern tail is
longer than the northern one and features a putative dwarf galaxy at
the tip \citep{Schw78, Mirabel92}.  Images in {\Ha} of NGC 4038/39 show
vigorous star formation within the main disks \citep{Whit95}.  The
optical spectrum, with a blue continuum and strong emission lines, is
indicative of an active star--forming galaxy.  The equivalent width of
the {\Ha} line is $2$--$3$ times higher than the normal average of
$\sim25$~{\AA} in an average Sb or Sc galaxy, categorizing NGC 4038/39
as a starburst galaxy \citep{Kennicutt92}.

The southern tail is rich in {\HI}, while the northern is {\HI}--poor.
{\HI} contours overlay the optical image in Figure~\ref{fig:overview}
\citep[from][]{hib99}.  At the end of the southern tail lie at least
three {\HII} regions.  A spectrum from \citet{Mirabel92} shows that
this region has a low O/H abundance, which is typical for dwarf
galaxies.  The region at the tip of the tail also shows a high
concentration of {\HI} \citep{vdH79, hib99}.

We use the maximum projected tail length of $65$~kpc \citep[corrected
for our adopted distance]{Schw78} and the disk rotation speed of
$150$~\kms inferred from the Fabry--Perot observations of
\citet[corrected for disk inclination]{Amram92}, to infer an
interaction age of $65~{\rm kpc} / 150~{\kms} = 420$~Myrs.  This age,
measured from orbital periapse when the tails were launched, is in
excellent agreement with the best match time of $450$~Myrs from the
numerical model of NGC 4038/39 run by \citet{Barnes88}.  It is also
remarkably close to the age of the population of $500$~Myr old
globular clusters found by \citet{whit99} \citep[see also][]{fritze98},
supporting the hypothesis that this globular cluster population formed
around the time when the tails were first ejected \citep{whit99}.

\subsubsection{Inner Region}
\label{sec:cen4038}

\citet{whit99} found young globular clusters in the inner regions of
the Antennae. More than $14,000$ point--like objects were identified
in the field, of which most were probably young stars formed in the
merger.  Depending on selection criteria, the number of young compact
star clusters is between $800$ and $8000$.  Remaining sources are
individual supergiant stars, also in the Antennae.  The luminosity
function is composed of two power law segments with a bend at $M_V\sim
-10.4$.  The median effective radius of the young clusters is $4 \pm
1$~pc, just slightly larger than the $3$~pc radius for Milky Way
globular clusters \citep{vdb96}.  The ages of the young clusters range
from less than $5$~Myr up to $100$~Myr. There are also populations of
intermediate age ($500$~Myr) and of old globular clusters
\citep{whit99}.

\subsubsection{Observations of the Southern Tail}
\label{sec:obs4038}

Three observations were planned for this merger, but only one has been
included in this study.  In Figure~\ref{fig:overview}, the WFPC2 field
of view is overlaid on an optical image of NGC 4038/39.  The WFPC2
image was obtained on 1999 Feb 22 (see the $V$--band image in
Figure~\ref{fig:V4038}).  Two later observations, not analyzed here,
are indicated by the lighter boxes in the image.  At the distance
modulus of $31.4$ \citep[for $H_0 = 75$~km/s/Mpc; as in][]{whit99},
the limiting magnitude ($50$\% completeness) of $m_V\sim25.8$,
consistent with our visual estimate of the completeness correction
from the color magnitude diagram (CMD), corresponds to an absolute
magnitude limit of $M_V\sim-5.8$.  Despite the closeness of this
merger, a long exposure was obtained (see Table~\ref{tab:obs}) to
match the depth of existing images of the inner regions. Therefore,
these NGC 4038 images are far deeper in absolute magnitude than
those of the other mergers presented in this paper.  In addition to
our image along the middle of the Southern tail, HST/WFPC2
observations of the candidate tidal dwarf galaxy at the tip of the
Southern tail were obtained by \citet{Saviane03}.  Their image is $2$
magnitudes deeper than our limiting magnitude.

\subsubsection{Results}
\label{sec:res4038}
\begin{figure}
\figurenum{9}
\plotone{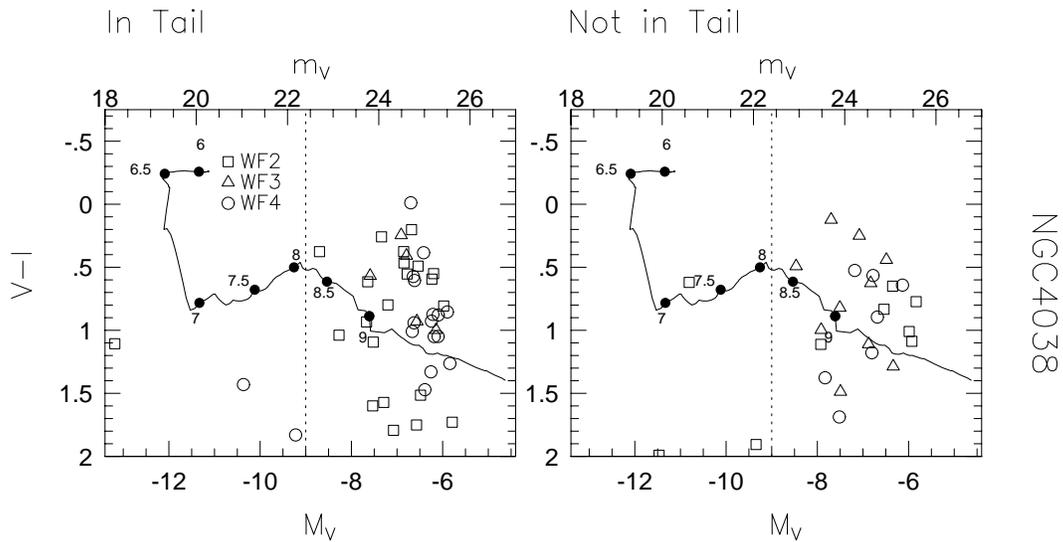}
\caption{
Color magnitude diagram for sources in and out of the tail region of
NGC 4038.  Color $V-I$ is plotted versus absolute magnitude $M_V$, on
the lower horizontal scale, and apparent magnitude $V$, on the upper
horizontal scale.  The absolute magnitude scale enables comparison
between different tails.  All sources, both in and outside of the
tail, with errors in $V$ less than $0.25$ magnitudes and with $-0.75 <
V-I < 2.0$, are included.  Sources coming from the different WF chips
are identified by special symbols as indicated in the legend.  No
corrections have been made for incompleteness.  The vertical dashed
line indicates a rough boundary such that brighter sources are more
likely to be clusters, rather than individual stars. The curve
represents the evolutionary track for an instantaneous burst model for
a $10^5$~{\msun} cluster with solar metallicity and a Salpeter IMF
with a 125~\msun cutoff, marked with log(age) in years \citep{bc93,
charlot96}.  This model would shift by a magnitude horizontally for
every factor of $2.5$ in mass.
\label{fig:cmd4038}}
\end{figure}

A $V-I$, $V$ color magnitude diagram (CMD) is presented in
Figure~\ref{fig:cmd4038}.  The plot only includes the range $-0.75 <
V-I < 2.0$ since we do not expect to see clusters with $V-I > 2.0$. 
In the inner region of NGC 4038/39, clusters identified by
\citet{whit99} have an $M_V < -9$ (indicated by the dotted line in
Figure~\ref{fig:cmd4038}) and colors $0.0 < V-I < 0.6$, with a
particular concentration at $-12 < M_V < -10$ and $0.3 < V-I < 0.4$
\citep{whit99}.  Of the sources in the tail, none lie in the same
region of the CMD where \citet{whit99} discover clusters, as shown in
Figure~\ref{fig:cmd4038}.  The tail sources are fainter and/or redder.
The grouping of $\sim30$ sources in the tail CMD falls in the same
area as individual stars in the inner region of the Antennae
\citep[$M_V > -9$ and $-1.0 < V-I < 1.0$;][]{whit99}.  Many of these
are probably isolated red or blue supergiants.  \citet{whit99} use $U$
and $B$ images to help separate stars from the cluster population,
however we have only $V$ and $I$ images.

\begin{figure}
\figurenum{10}
\plotone{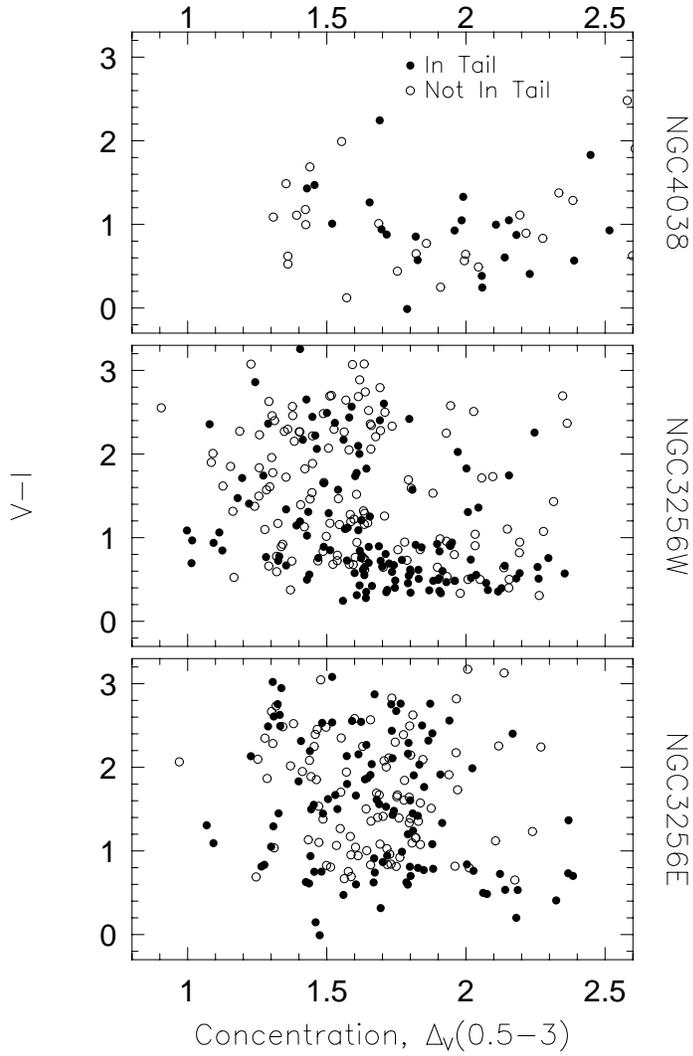}
\caption{
Concentration index plots for the two nearest mergers, NGC 4038 and
NGC 3256
(Western and Eastern tails). 
Color $V-I$ is plotted versus concentration index $\Delta_V(0.5-3)$.
The concentration index (a rough measure of cluster size) was
calculated from the difference in $V$ magnitude between an aperture of
$0.5$ and $3.0$ pixel radii.  Sources within the tail are indicated by
solid circles, while those outside the tail are indicated by open
circles.
\label{fig:colorcon}}
\end{figure}

Since the ``in--tail'' and ``out--of--tail'' regions in this field are
equal in area, it is noteworthy that there are a larger number of
faint sources in the tail regions ($42$ vs. $24$ in the
``out--of--tail'' regions).  However, there is no significant
difference between the distribution of sources in the CMD for sources
in and out of the tail (Figure~\ref{fig:cmd4038}).  The probabilities
of the magnitudes and colors being drawn from the same distributions
are $0.27$ and $0.89$, respectively, by the Kolmogorov--Smirnov test
\citep{numrec}.  In Figure~\ref{fig:colorcon}, the $V-I$ color is
plotted versus the concentration index, but again there is no
difference in the distributions for sources in and out of the tail
($P(KS) = 0.89$ for the color and $0.71$ for the concentration index
distribution).

\subsubsection{Discussion}
\label{sec:dis4038}
As seen in Figure~\ref{fig:V4038}, some of the fainter point sources
lie along the optical tail.  However, there is an absence of bright
($M_V < -9$), young clusters compared to the inner region.  If
clusters formed only concurrent with the tail, $\sim450$~Myrs ago,
they should still be detected in the CMD (see
Figure~\ref{fig:cmd4038}).  This is evident since $21$ $M_V < -9$
sources of intermediate ages were found by \citet{whit99} in their
comparable exposure of the body of the NGC 4038/39 merger and its inner
tail regions.  In fact, many of the fainter sources that we find
within the tail may be individual red and blue supergiants, perhaps
forming in the tail.

This image was taken $\sim30$~kpc from the inner region and contains
$\sim20$~kpc of the Southern tail.  It is possible that other regions
of the tails do have cluster populations.  \citet{whit99} identified
about a dozen objects in the inner part of the Southeastern tail, but
only one of them was bright enough ($M_V = -9.48$) to be definitively
identified as a cluster.  Our additional observations include the
Northern tail, which has less {\HI} by a factor of five \citep{vdH79,
hib99}, and another region in the Southern tail which is closer to the
inner region.  These images provide the opportunity to study the
distribution of clusters in the tail and the relationship of the {\HI}
content to the characteristics of the cluster population.  The
candidate tidal dwarf galaxy at the tip of the Southern tail has $8$
extended young star clusters, $5$ of which are brighter than $M_V =
-8.5$ and many young blue stars fainter than $M_V = -8.5$
\citep{Saviane03}.  At least the tip of the Southern tail of NGC
4038/39 contains star clusters.

\subsection{NGC 3256}
\label{sec:s3256}
NGC 3256 is intermediate in the ``Toomre Sequence,'' and it is the
second nearest merger in this study.  The inner region has structures
such as loops, knots, and dust lanes usually seen in mergers, as well
as sweeping, symmetric tidal tails (see the optical image in
Figure~\ref{fig:overview}).  In radio continuum observations, two
nuclei can be seen, separated by $\sim1$~kpc \citep{norris95}.  Of the
``Toomre Sequence'' mergers, NGC 3256 has the most molecular gas, $1.5
\times 10^{10} M_{\sun}$ \citep{Casoli91, Aalto91, Mirabel90}, and is
the brightest in the far--infrared, with $L_{FIR} = 3 \times 10^{11}
L_{\sun}$ \citep{sargent89}.  In addition, NGC 3256 is the most X--ray
luminous starburst known, with $L_{0.5-10\rm{keV}} = 1.6
\times10^{42}~\rm{ergs~s}^{-1}$ \citep{Moran99}.  The tails have rich
{\HI} content but no tidal dwarf candidates have been discovered
within them \citep[see the {\HI} contours in
Figure~\ref{fig:overview}]{English}.

The approximate dynamical age of NGC 3256 is calculated in a similar
manner as for NGC 4038/39 (presented in \S\ref{sec:s4038}).  However,
for a face--on orientation the maximum velocity is best represented by
the escape velocity, $150$~{\kms} \citep{English00}, and the tail age
is estimated to be $61~{\rm kpc}/150~{\kms} = 400$~Myr.

\subsubsection{Inner Region}
\label{sec:cen3256}

In WFPC2 images, \citet{zepf99} find more than $1000$ compact, bright
blue objects in the inner regions of NGC 3256.  This population has
colors, luminosities, and sizes similar to those of young globular
clusters, with ages from a few to several hundred Myr.  The luminosity
function can be fit with a power law.  The young cluster population is
quite extreme in that it comprises $\sim20$\% of the total blue
luminosity of the galaxy.  This could indicate a high global
efficiency of cluster formation.

\subsubsection{Observations of the Western and Eastern Tails}
\label{sec:obs3256}

In Figure~\ref{fig:overview}, the WFPC2 field of view is overlaid on
an optical image of NGC 3256.  The $V$ and $I$--band images of the
Western tail were obtained on 1999 Mar 24, and we present the
$V$--band image in Figure~\ref{fig:WV3256}.  The images of the Eastern
tail, obtained on 1999 Oct 11, are represented by the $V$--band image
in Figure~\ref{fig:EV3256}.  The $50$\% and $90$\% completeness limits
correspond to $M_V=-7.5$ and $M_V=-8.7$.

\subsubsection{Results}
\label{sec:res3256}
The observations of the inner region from \citet{zepf99} were obtained
in the $B$ and $I$ bands, which precludes a direct comparison with our
observations in $V$ and $I$.  From the $V$--band WFPC2 images of the
Western and Eastern tails in Figures~\ref{fig:WV3256} and
\ref{fig:EV3256}, it can clearly be seen that the density of objects
is higher within the tail regions.  In addition, both the Western and
Eastern tails show a grouping of sources in the CMD at $-10 < M_V <
-7$ and $0 < V-I < 1$, which is more prominent in the ``in--tail''
than in the ``out--of--tail'' regions (see Figure~\ref{fig:cmd3256}).
For the Western tail, the Kolmogorov--Smirnov (K--S) statistic
indicated that the probabilities that the magnitude and colors of
``in--tail'' and ``out--of--tail'' sources are drawn from the same
population are, $0.03$ and $2.2\times10^{-6}$, respectively.  For the
Eastern tail, the statistical significance of the difference is not as
large.  The magnitudes have a K--S probability of $0.20$ of being
drawn from the same distribution, but the color distributions are
different at $> 2\sigma$ significance.  Some of the fainter sources
could be individual stars, but the majority are likely to be clusters
\citep{whit99}.  In the Western tail, there are $71$ sources falling
in the region of the CMD defined by $-10 < M_V < -7$ and $0 < V-I <
1$.  The Eastern tail has $25$ sources in this region.  Outside the
tail, there are only $29$ sources in the Western image and $8$ in the
Eastern image that fall within this same region of the CMD.

The cluster candidate sample is contaminated by foreground stars since
NGC 3256 is at a low galactic latitude ($b = +11.7$).  Examination of the
concentration index can aid in separating the clusters from the
stars. The concentration index for blue sources ($V-I < 1.1$) within
the Western tail (Figure~\ref{fig:colorcon}) is peaked at $\sim1.8$,
while for red sources ($V-I > 1.1$) out of the tail, the peak is
$\sim1.5$.  Hence the clusters (blue objects) appear to be resolved
compared to the stars (red objects).  By the K--S test, these
distributions have a probability of $0.06$ of being drawn from the
same distribution of concentration index, but the distributions of
colors differ at $> 5\sigma$.  There is a grouping of sources in the
concentration index plot at a large concentration index which is
composed of the same sources as the grouping at $0 < V-I < 1$ in the
CMD.

Candidate cluster sources in the Eastern tail are less numerous than
those in the Western tail, but many of them are in the same region of
the CMD and the concentration index plot as those Western sources that
have been identified as clusters.  In the Western tail, there is a
clear separation in concentration index between ``in--tail'' and
``out--of--tail'' sources.  The sources in the Eastern tail have less
separation, no significant difference in the distribution of
concentration index by the K--S statistic ($P(KS) = 0.84$), and a
$2\sigma$ difference in the distributions of color.

The majority of clusters in the Eastern tail, $18$ out of $25$,
appear in WF2, which includes the part of the tail closest to the
center.  In the Western tail, the majority of clusters, $43$ out of
$71$, appear in WF4, compared to $25$ in WF2, the other significant
contributor.  In this case, WF4 contains the part of the tail farthest
from the center.

\subsubsection{Discussion}
\label{sec:disc3256}

\begin{figure}
\figurenum{11}
\plotone{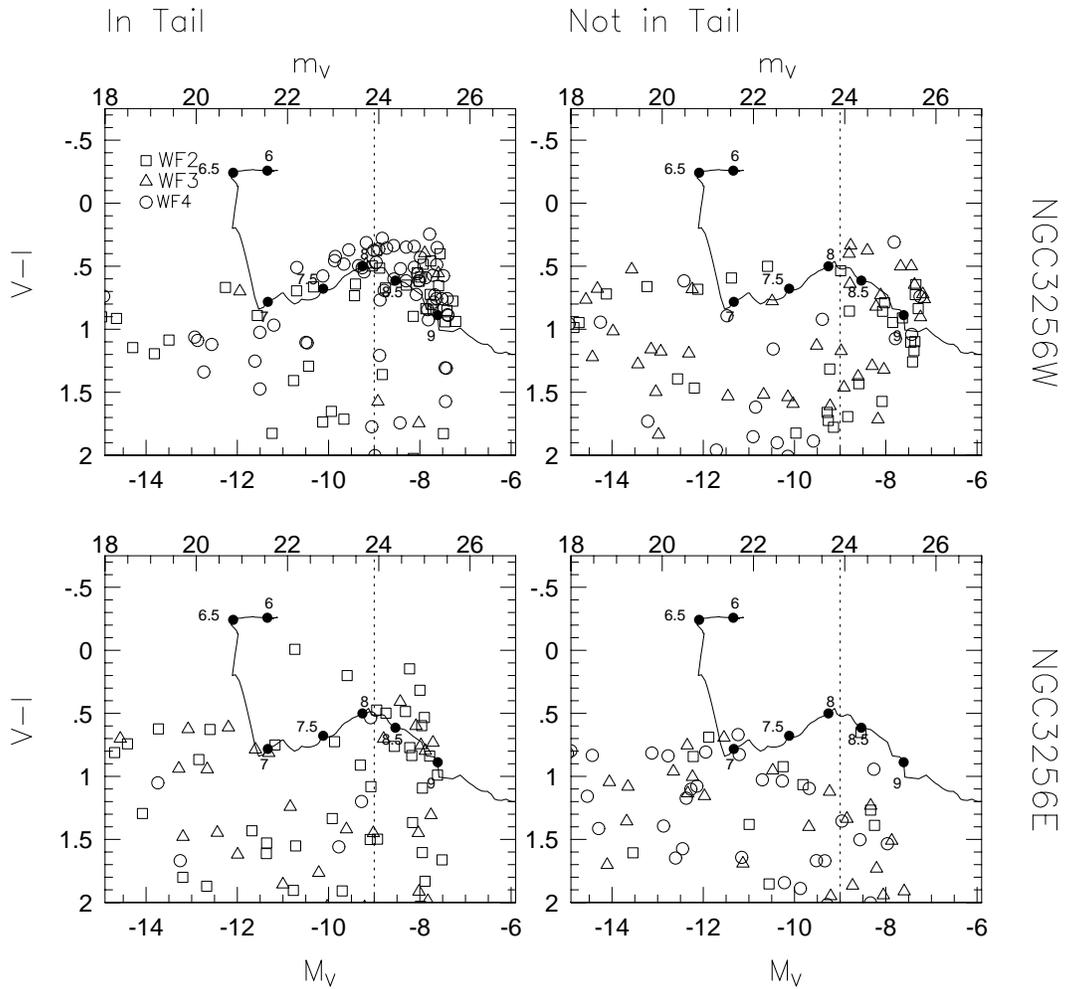}
\caption{
Color magnitude diagrams, same as Figure~\ref{fig:cmd4038}, but for
NGC 3256.  The Western and Eastern sources, in and out of the tails,
are given in the top and bottom rows, respectively.
\label{fig:cmd3256}}
\end{figure}

From the CMD (Figure~\ref{fig:cmd3256}) and the concentration index
diagram (Figure~\ref{fig:colorcon}) for the Western tail, we infer
that most of the sources with $-10 < M_V < -7$ and $0 < V-I < 1$ are
star clusters.  The peak of the color distribution is $V-I \sim 0.5$.
The majority of clusters in the Western tail are in the outer higher
surface brightness regions of the tail.

There are fewer clusters in the Eastern tail and the majority of them
are closer to the inner regions.  The Eastern tail appears to be of
higher surface brightness toward the center, following the
distribution of clusters.  In addition to more star clusters, the
Western tail has a a higher {\HI} mass than the Eastern tail
\citep[$2.2\times 10^{9}~$\msun vs. $1.4\times 10^{9}$~\msun;][]{English}.

Ages of clusters can be estimated using the evolutionary tracks for a
$10^5~$\msun instantaneous burst \citep{bc93, charlot96}
\footnote{Taken from
http://www.stsci.edu/instruments/observatory/cdbs/cdbs\_bc95.html}
with solar metallicity and a Salpeter IMF, superimposed on the CMDs in
Figure~\ref{fig:cmd3256}.  A range of cluster masses, $\sim 10^4$ --
$10^5~$\msun, would allow for a population with the range of ages,
$30$--$300$ million years, to explain most of the observed $M_V$ and
$V-I$.  However, the bluer clusters (those with $V-I < 0.3$) require
either a lower metallicity \citep[see][]{setal96} or considerably
younger populations (several million years old).  Regardless, it would
appear that the clusters are younger than the tail in which they
formed.  Therefore, they were not pulled out into the tail from inner
regions, but rather they formed ``in situ'' in dense {\HI} regions
within the tail.

\subsection{NGC 3921}
\label{sec:s3921}

NGC 3921, the second to last of the ``Toomre Sequence'', features a
single nucleus, a main body with ripples and loops, and a pair of
crossed tidal tails.  The two galaxies have essentially merged
\citep[Figure~\ref{fig:overview}; see also][]{S96}.  The southern tail
extends below the main body, ending in large concentration of {\HI},
possibly a tidal dwarf galaxy \citep{HvG96}.  NGC 3921 is the most
distant merger in our sample of four.

Consistent with an evolved merger, the optical spectrum is dominated
by Balmer absorption lines, a classic post--starburst signature
\citep{Kennicutt92}.  NGC 3921 shows no {\HI} in the northern tail
while the southern tail is gas--rich (see the {\HI} contours in
Figure~\ref{fig:overview}).  This difference in {\HI} between the
tails may indicate the merger of two different types of galaxies, a
gas--rich, late--spiral type and a gas--poor, early--type disk galaxy.
The approximate tail age of NGC 3921, calculated as in
\S\ref{sec:s4038}, is $64~{\rm kpc}/140~{\kms} = 460~{\rm Myr}$
\citep{HvG96}.

\subsubsection{Inner Region}
\label{sec:cen3921}

In WFPC2 images obtained by \citet[hereafter S96]{setal96}, there are
two chains of point-like candidate globular clusters and more extended
stellar associations along the inner part of the southern tail, as
well as several slender, narrow arcs of these sources.  Their HST
image reveals $\sim 100$ candidate globular clusters and $\sim 50$
stellar associations.  The globular clusters tend to be more centrally
located while the associations trace the southern loop and two tidal
tails.

The majority of the inner region's clusters and associations are blue
and have rather uniform colors with a median $V-I = 0.65$ for globular
clusters and $0.56$ for associations.  Depending on the metallicity,
the ages of the globular clusters range from $250$--$750$~Myr with an
upper limit on their radii of $5$~pc.  The luminosity function of the
globular clusters and that of the associations are represented by
power laws.  S96 estimate that the number of globular clusters
increased by greater than $40\%$ during the merger.

\subsubsection{Observations of Southern Tail}
\label{sec:ob3921}

In Figure~\ref{fig:overview}, the WFPC2 field of view is overlaid on
an optical image of NGC 3921 (see Table~\ref{tab:obs} for exposure
information).  The images of the Southern tail, obtained on 1999 Apr
30, are represented by the $V$--band image in Figure~\ref{fig:V3921}.
With a distance modulus of $34.5$ (for $H_0 = 75$~km/s/Mpc; S96), our
source catalog for NGC 3921 is $50$\% complete to $m_V \sim 25.8$,
consistent with our visual estimate of the completeness correction
from the CMD, which at this large distance corresponds only to $M_V =
-8.7$.

\subsubsection{Results}
\label{sec:res3921}

Because our image partially overlaps with the observation by S96, we
compare three globular cluster and five stellar association candidates
found in that paper which correspond to our sources. These eight
sources have a range in brightness of $23.5 < V < 25.4$ and a range in
color of $-0.2 < V-I < 1.1$.  The mean difference between the S96 data
and our data is $0.12$ magnitudes for $V-I$ and $0.43$ magnitudes for
$V$.  The S96 values are consistently fainter than our values and
slightly redder on average.  The differences between the data appear
to be due to the differences in the corrections for finite aperture
size, charge-transfer efficiency (CTE), and reddening.  For sources on
the WF chips, S96 used a $1$ pixel aperture for aperture photometry,
while our procedure used a $2$ pixel aperture.  This would cause more
extended sources, or stellar associations, to have a larger difference
between the two data sets than the less extended sources since a
single aperture correction has been made for all sources.  In our
comparison, the five associations have a mean difference of $0.55$ mag
in $V$ and $0.12$ mag in $V-I$ while the cluster candidates have a
mean difference of $0.23$ mag in $V$ and $0.09$ mag in $V-I$.
Corrections for CTE problems were not made by S96 due to the varied
background in their images.  Our calculations include corrections for
CTE, at times as large as $-0.2$ magnitudes for faint sources near the
top of the chip.  S96 used $A_V = 0.00$ to correct for foreground
Milky Way extinction, whereas we used $A_V = 0.046$.  As an
independent check on their photometry, S96 compared their photometry
performed with VISTA software with photometry using IRAF software.
The objects measured in IRAF were found to have a mean difference in
$V-I$ of $0.08$ mag bluer than the objects measured in VISTA, possibly
due to different centering algorithms or different aperture
corrections.  Regarding $V$ magnitudes, S96 estimate that larger
associations may have $V$ magnitudes up to several tenths of a
magnitude brighter, due to their more extended profile. Hence, taking
into account differences in corrections for aperture size, CTE, and
reddening, we find our values for photometry to be consistent with the
S96 photometry.

\begin{figure}
\figurenum{12}
\plotone{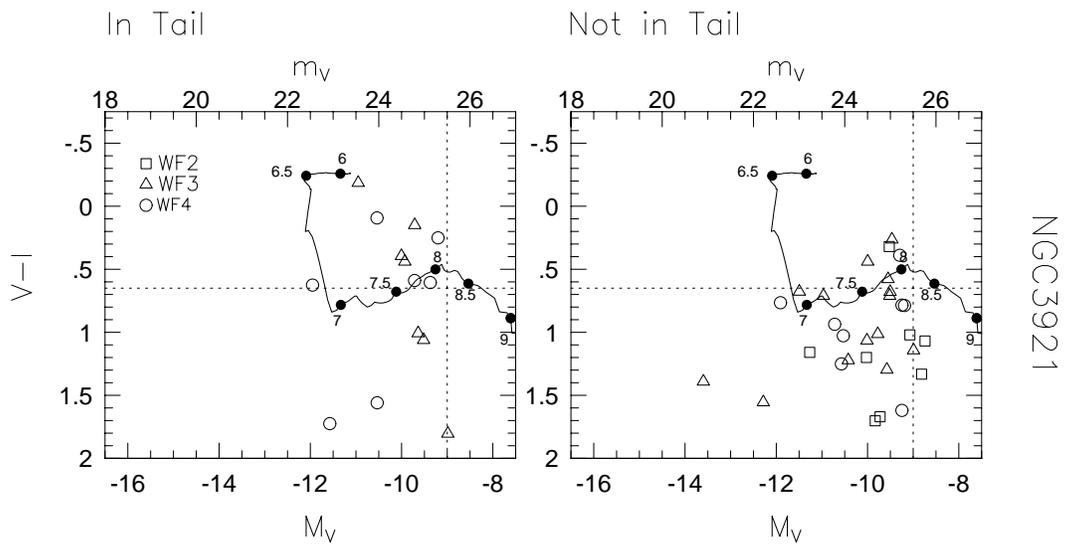}
\caption{
Same as Figure~\ref{fig:cmd4038}, but for NGC 3921.  The horizontal
line represents the median $V-I$ for the clusters in the inner region
of the merger.
\label{fig:cmd3921}}
\end{figure}

As shown in Figure \ref{fig:cmd3921}, there are few sources seen in
the tail of this system.  S96 find a population of bright clusters in
the inner region with a narrow color range at $V-I = 0.65$, indicated
by the horizontal dotted line in Figure~\ref{fig:cmd3921}.  Sources in
the tail do not appear to be clustered around this line, indicating
that there is not a large population of clusters of the same age in
the tail.  The K--S test shows no significant difference between the
distributions of color and magnitude ($P(KS) = 0.98$ and $0.05$) for
sources in the tail compared to those out of the tail.

The associations that overlap with S96 fall in the same region of the
CMD, $-11 < M_V < -9$ and $-0.1 < V-I < 0.9$.  There are a few
additional sources in the tail that also fall in this region of the
CMD, and most of these lie closer to the inner region of the merger.

In a first pass through the photometry, only point sources were kept,
using the method outlined in \S\ref{sec:phot}.  This method yielded
very few sources and missed several of the associations identified in
S96.  When the criteria were relaxed, as outlined in
\S\ref{sec:detect}, many more extended objects were found.  Most of
those found off the tail may be background galaxies, but those within
the tail have properties consistent with the stellar associations
found in S96.  At this distance one WF pixel corresponds to $39$~pc,
so it is difficult to distinguish clusters from stars using the
concentration index.

The putative tidal dwarf galaxy was imaged in the PC.  It appears to
be a low surface brightness dwarf and no point sources were found
within it.  However, there are four knots within the more diffuse
emission which extends over about $7$~kpc.  This tidal dwarf galaxy
candidate will be addressed in a future study of the extended sources
within these tails.

\subsubsection{Discussion}
\label{sec:disc3921}

Most of our sources that are in the same region of the CMD as those
from S96 are located in the region closer to the main body of the
merger and are bluer than sources elsewhere in the field.  Most of the
bluer sources in the tail have a large concentration index, indicating
the possible presence of larger, stellar associations.  In addition to
these large associations, there is a putative dwarf galaxy centered in
the PC.  There is an absence of point sources in the optical tail at
larger distances from the main body of the merger.

\subsection{NGC 7252 (``Atoms for Peace'')}
\label{sec:s7252}
As the system located at the end of the Toomre Sequence, NGC 7252
exemplifies the prototypical merger remnant (see the optical image in
Figure~\ref{fig:overview}).  It is the second most distant of the four
mergers in our sample.  \citet{Schw78, S98} defines some
characteristics of a recent merger remnant: a pair of long tidal
tails, isolated from neighbors, a single nucleus, tails moving in
opposite directions relative to nucleus, and chaotic motions in the
main body of the merger, all of which NGC 7252 possesses
\citep{Schw82}.  The central region of this remnant has a single
nucleus with several loops.  Observations with HST have revealed a
mini--spiral structure in the inner region which coincides with the
disk of molecular and ionized gas \citep{Wang92, Whit93, miller97}.
The long, straight tails extend from the body of the remnant to the
east and northwest, each ending in a concentration of {\HI} associated
with a candidate tidal dwarf galaxy.  The outer parts of both tails
have blue $B-R$ colors, with the bluest parts coincident with the
highest gas density \citep{hib94}.
 
The tails of NGC 7252 and the western loop region are both rich in
{\HI} (see the {\HI} contours in Figure~\ref{fig:overview}).  The
kinematics of the northwestern tail suggest that the material at the
base of the tail is falling into the main remnant \citep{HM95}.  This
infall could explain some of the peculiar ripples and shells
associated with the center.  The presence of a young, metal poor
emission line cluster in the western loop ($15$~kpc from the center),
is consistent with recent star formation that is perhaps triggered by
infalling metal poor gas from the tail \citep{ss98}.  The approximate
tail age of NGC 7252, calculated as in \S\ref{sec:s4038}, is $160~{\rm
kpc} / 220~{\kms} = 730~{\rm Myr}$ \citep{hib94} which is in good
agreement with the age of 770 Myrs as determined through numerical
modeling \citep[correcting to $H_0 = 75$~km/s/Mpc]{HM95}.

\subsubsection{Inner Region}
\label{sec:cen7252}

WFPC2 observations of the inner regions of NGC 7252 detected $\sim500$
cluster candidates which separate into three populations.  Luminous,
blue clusters with a narrow color range $V-I \sim 0.65$, have ages
around $650$--$750$~Myr.  They were likely to have formed around the
time that the tails were launched.  These clusters have an upper size
limit of $4.8\pm0.4$~pc.  The inner disk is home to a population of
very young ($\sim10$~Myr) clusters with $U-B$ colors indicating that O
stars dominate and with radii of $8.3\pm0.6$~pc.  The third population
consists of the older, metal--poor globular clusters from the original
galaxies.  The combined luminosity function of these populations of
clusters is a power law with $\alpha \sim -1.8$ \citep{miller97}.

\subsubsection{Observations of Western and Eastern Tails}
\label{sec:ob7252}
In Figure~\ref{fig:overview}, the WFPC2 fields of view are overlaid on
an optical image of NGC 7252 (see Table~\ref{tab:obs} for exposure
information).  The images of the Western tail, obtained on 1998 Nov
18, are represented by the $V$--band image in Figure~\ref{fig:WV7252}.
The Eastern tail observations, obtained on 1999 Aug 29, are
represented by the $V$--band image in Figure~\ref{fig:EV7252}.  With a
distance modulus of $34.0$ \citep[for $H_0 = 75$~km/s/Mpc;][]{Schw82},
the detection limit of $m_V \sim 26.0$ ($50$\% completeness),
consistent with our visual estimate of the completeness correction
from the CMDs, corresponds to $M_V = -8.1$.

\subsubsection{Results}
\label{sec:res7252}
\begin{figure}
\figurenum{13}
\plotone{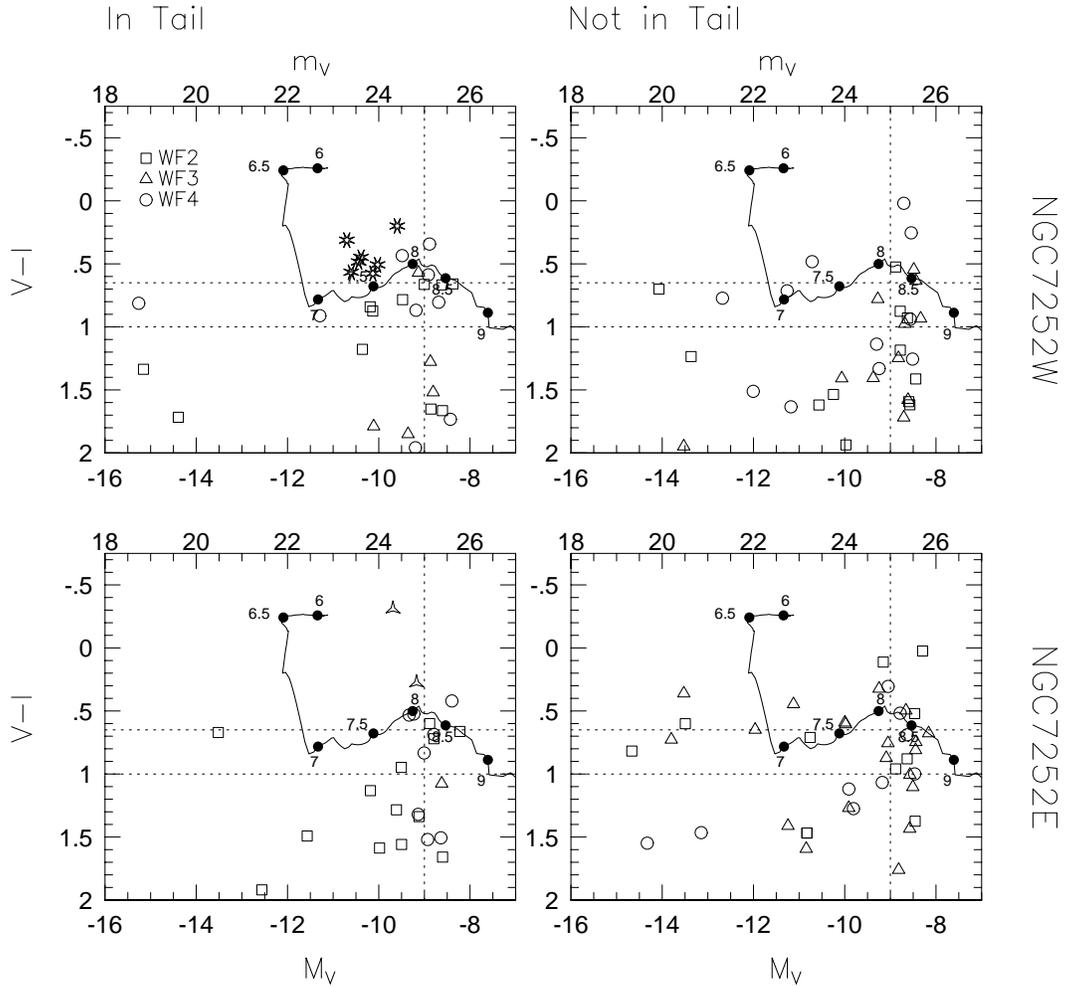}
\caption{
Same as Figure~\ref{fig:cmd4038}, but for NGC 7252.  The Western and
Eastern sources, in and out of the tails, are given in the top and
bottom rows, respectively.  The sources in the tidal dwarfs at the end
of the Western and Eastern tails (see Figure~\ref{fig:dw7252}) are
indicated as ``star symbols'' of the shape corresponding to the
appropriate WF chip.  For comparison, the horizontal dotted lines
represent the colors of the two main populations of clusters in the
inner region of this merger.  The evolutionary track for a single
burst Bruzual and Charlot model \citep{bc93} is given as in previous
CMDs.
\label{fig:cmd7252}}
\end{figure}

The clusters in the inner region identified by \citet{miller97}
include populations of intermediate and old clusters at $V-I=0.65$ and
$1.0$, as indicated by the horizontal dotted lines on the CMD in
Figure~\ref{fig:cmd7252}.  The blue clusters have $19 < V <25$.  There
are also some bright, blue clusters with $V-I < 0.6$ \citep{miller97},
found mostly within the inner $6${\arcsec} of the galaxy.

\begin{figure}
\figurenum{14}
\plotone{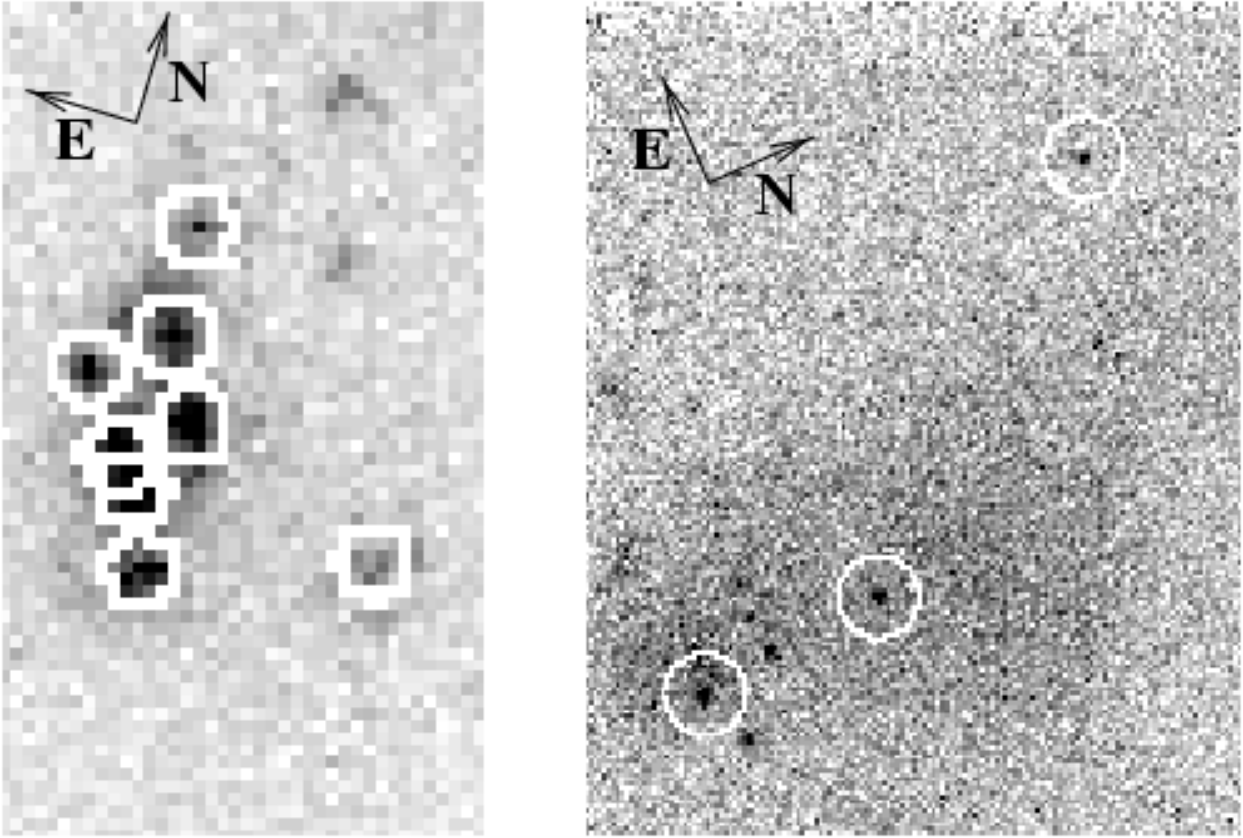}
\caption{
The dwarf galaxies at the ends of the Western (left) and
Eastern (right) tails of NGC 7252, taken in the
F555W ($V$) filter.  The colors and magnitudes of several
clusters within these dwarfs are shown as ``star symbols''
in the CMDs of Figure~\ref{fig:cmd7252}.
The clusters in the Western dwarf are found to be relatively blue
($V-I\sim0.4$), somewhat younger than the dominant population in the 
inner region which has an age of $\sim650$--$750$~Myr.
Those in the Eastern tail are extremely blue ($V-I < 0.3$),
consistent with the youngest population in the inner regions,
having an age of $\sim10$~Myr.
\label{fig:dw7252}}
\end{figure}

The Western tail CMD shows a possible enhancement of bluer, faint
sources at $-12 < M_V < -8.5$ and $0.2 < V-I < 1.0$
(Figure~\ref{fig:cmd7252}), which are located in the tail region.  The
K--S test yields a difference at a $2\sigma$ level between the colors
for ``in--tail'' and ``out--of--tail'' sources, but no significant
difference in their magnitude distributions.  Seven members of the
group at $0.2 < V-I < 0.65$, noted as star symbols on the CMD, are
located in the putative tidal dwarf galaxy.  A close up of the V-band
image of this region with the seven star clusters indicated is shown
in Figure~\ref{fig:dw7252}.  The several sources with $0.65 < V-I <
1.0$ are spread through the optical tail.

For the Eastern tail, Figure~\ref{fig:cmd7252} shows a small group of
sources in the tail at $M_V\sim-9$ and $0.2 < V-I < 1.2$. However,
this concentration is not different from the grouping in the CMD of
sources not in the tail.  There is only one bright candidate ($M_V <
-9$) in the tail region, while the region outside the tail hosts
several such bright sources at $V-I\sim0.5$.  The nature of these
sources is ambiguous.  The K--S statistic shows no significant
difference in either the magnitude or the color distributions for
``in--tail'' and ``out--of--tail'' sources.  The concentration index
is less useful for this merger because it is so far away ($1$ WF pixel
is $~31$~pc).  There are two sources in the Eastern tail dwarf which
are quite blue ($V-I < 0.3$), noted as three--pointed star symbols on
Figure~\ref{fig:cmd7252}; see Figure~\ref{fig:dw7252} for a $V$--band
image of the dwarf.

\subsubsection{Discussion}
\label{sec:dis7252}
Although there are several blue sources with $V-I < 1.0$ in the
Eastern and Western tail regions, there is no convincing evidence for
an excess over the ``out--of--tail'' regions, with the important
exception of clusters within the prominent tidal dwarf candidates in
each tail.

The tidal dwarf candidate in the Western tail is clumpy, consisting of
several knots spread over $\sim1$~kpc and embedded in extended high
surface brightness material.  There are seven clusters found within it
and they are relatively blue.  Their concentration indices are
somewhat larger ($2.2$--$3.0$) than the mean of $1.8$ for the tidal
cluster sources, but they are still consistent with being clusters at
this distance.  However, the measurement of the concentration
parameters for clusters in this tidal dwarf candidate is uncertain
since it is a crowded environment.  These clusters in the Western
tidal dwarf candidate are grouped together in the CMD (star symbols in
Figure~\ref{fig:cmd7252}) and their average $V-I$ color is $0.4$,
bluer than the major population of clusters in the inner regions of
NGC 7252.  The CMD suggests an age of less than $100$~Myrs, which is
less than the formation age of the tail, and as such the clusters
formed within the tidal debris.

\citet{hib94} found strong kinematic evidence for a mass
concentration at the location of the putative tidal dwarf in the
western tail of NGC 7252.  This evidence consisted of an increase in
the {\HI} linewidth centered on and symmetric with the concentration
of gas, light, and star forming regions associated with the tidal
dwarf galaxy.  The appearance of star clusters at this location,
especially since there is no statistically significant excess of star
clusters elsewhere in the tail, is further evidence in support of this
region being dynamically distinct.

The tidal dwarf in the Eastern tail is almost featureless and of only
slightly higher surface brightness than the tail.  The several clusters 
within the
dwarf are among the bluest in the sample.  They are consistent with
the very blue colors of those clusters within the inner $6${\arcsec}
of NGC 7252.  These inner region clusters have an age of $\sim10$~Myr.
Though the clusters could have formed recently, the dwarf itself may
have formed earlier.

The contrast in the distribution of young stars in the two candidate
dwarf galaxies is striking.  The Western tail dwarf is of higher
surface brightness and contains several blue clusters in close
proximity, and the Eastern tail dwarf is almost featureless, with only
two clusters spaced farther apart.  In the Eastern tail dwarf, the
extreme blue colors indicate that the clusters are considerably
younger than the age of the interaction. A similar situation is found
in the Northern Starburst region of Stephan's Quintet, one of
Hickson's compact galaxy groups.  That region is separated from the
nearest giant galaxy by more than $25$~kpc, yet it hosts a very young
stellar population \citep[less than $5$~Myr;][]{sarah}.  It will be
interesting to compare the range of properties of these and other
tidally formed dwarfs to the varied population of compact and diffuse
dwarf galaxies in the Local Group.

%%%%%%%%%%%%%%%%%%%%%%%%%%%%%%%%%%%%%%%%%%%%%%%%%%%%%%%%%%%%%%%%%%%%%%

\section{Summary and Discussion}
\label{sec:conc}

In our study of six HST WFPC2 images of six tidal tails in four
merging pairs, we found evidence for extreme differences in cluster
formation among the tidal tails.  To compare the debris in the
different mergers on equal footing, we must consider contamination by
foreground and background objects and the relative areas of
``in--tail'' and ``out--of--tail'' regions (fractions $f_{in}$ and
$f_{out}$).  We must also consider selection effects due to the
different distances of the pairs.  Table~\ref{tab:summary} gives the
``in--tail'' and ``out--of--tail'' source densities in each tidal
tail, for sources with $V-I < 0.7$ and with $M_V < -8.5$.  The latter
criterion was chosen because sources this bright are likely to be
clusters rather than individual stars.  Table~\ref{tab:summary} also
lists the specific frequencies of young clusters, $S_{young}$, for
each tail region, defined as the number of young clusters per $M_V =
-15$ luminosity (in a similar way to how \citet{harris91} defined the
specific frequency for old globular clusters).

\begin{figure}
\figurenum{15}
\plotone{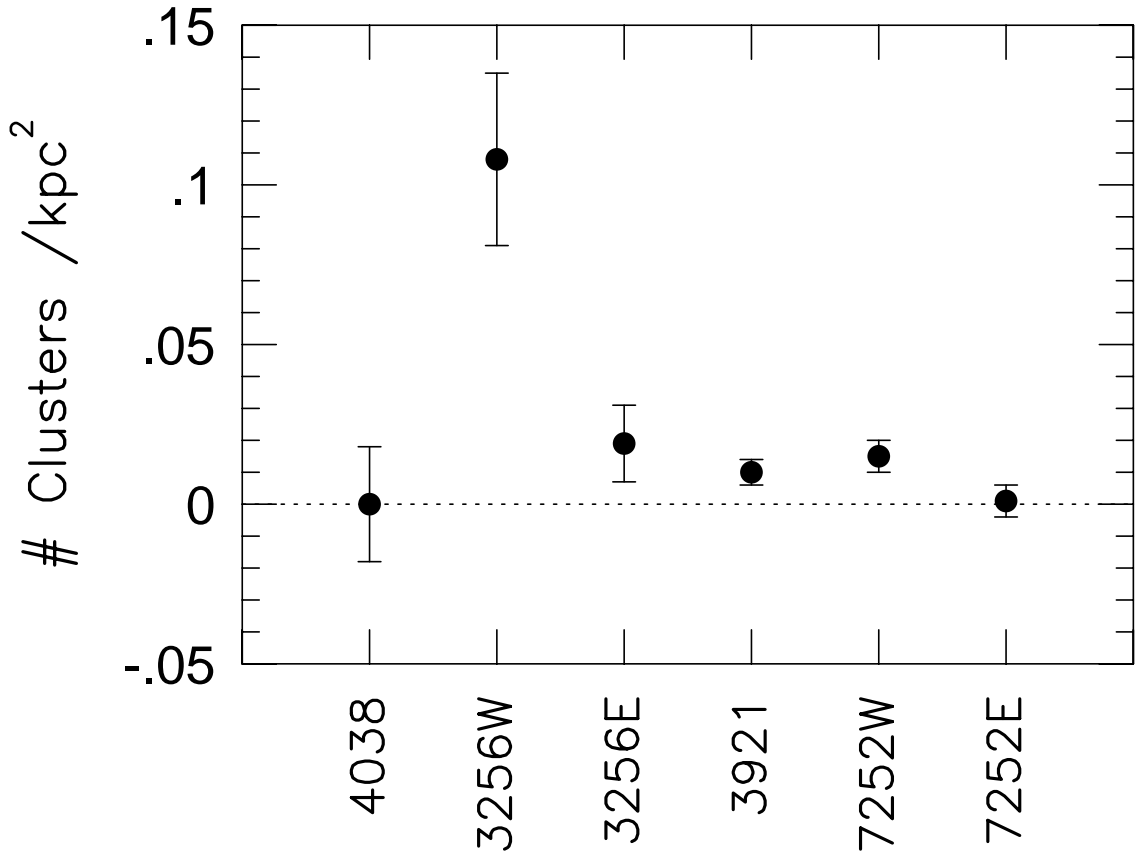}
\caption{
The number of candidate clusters per square kpc is 
compared for the six different areas of debris.  Background sources
were subtracted, statistically, using the source densities
in areas ``out--of--tail''.  The excesses ``in--tail'' are plotted.
Details of the calculation are given in Table~\ref{tab:summary}.
\label{fig:summary}}
\end{figure}

Figure~\ref{fig:summary} gives, for the six regions, the differences
in the source densities in and out of the tails.  This is a way of
subtracting the background, but it neglects the fact that some
``out--of--tail'' sources might be real clusters.  Errorbars are
simply from Poisson statistics; if some of the objects outside of the
tails are clusters then the background subtraction is systematically
too large and the surplus found would represent a lower limit.

The Western tail region of NGC 3256 shows a significant excess of
sources likely to be star clusters.  The five other regions show a
much smaller number, consistent with $0$ for NGC 4038 and NGC 7252E,
but statistically significant for the other three tail regions, NGC
3256E, NGC 3921, and NGC 7252W.  In the case of NGC 3921, many of the
sources in the Southern tail are stellar associations also reported by
S96.  By having different detection criteria for NGC 3921 and NGC
7252, we attempted to find any faint clusters that might add to the
numbers such that these regions might compete with the number of
clusters found in NGC 3256W.  Since we did not find a population of
faint sources in the tails of NGC 3921 and NGC 7252, the excess of
sources in NGC 3256W is even more significant.  If we just look at the
brightest sources in all the tails which have $M_V < -9.0$ and $M_V <
-9.5$ (with $V-I < 0.7$, as above), NGC 3256W still has a much larger
excess of clusters within the tail.

While the the tails of NGC 7252 generally lack clusters, it is
remarkable that there is a population of clusters associated with both
tidal dwarf candidates (seven in the western dwarf, two in the eastern
one).  The two tidal dwarf candidates show different populations of
clusters, however, with the Eastern dwarf bluer than the Western
dwarf. For the Western tidal dwarf, the clusters may have formed
either before or concurrent with the dwarf.  The Eastern dwarf
probably formed prior to its recent burst of star formation.  Though
differential reddening could induce the observed color differences
between the two dwarfs, we have no evidence which suggests this might
be the case.

The NGC 4038 tidal debris shows no large young cluster population
like that observed in its inner regions.  There are fainter objects,
which could be individual young stars within the tail, but there are
only $3$ objects brighter than $M_V = -9$, the cutoff for clusters
used by \citet{whit99}, and all of these have $V-I > 1$.  At least
this section of the tail, if not the entire tail, has few, if any,
clusters.  Interestingly, HST/WFPC2 images of the region of the tail
coincident with the tidal dwarf candidate \citep{Saviane03} show $8$
young stellar associations in the vicinity of the tidal dwarf. The
high concentration of blue star clusters in the vicinity of a tidal
dwarf candidate, but general lack of such clusters elsewhere in the
tail, is very similar to the situation for NGC 7252. It will be very
interesting to see if the other two unreduced pointings on the NGC
4038/39 tails (see Figure~\ref{fig:overview}) show a similar lack of
bright clusters.

Clearly, the Western tail of NGC 3256 has the largest population of
tail clusters observed in our sample.  Both the colors and
concentration of these objects are coincident with inner region
clusters in the merger.  The Eastern tail has a small excess ($\sim
2\sigma$) of clusters as well.

So why might the tails of NGC 3256 be a preferred environment for
extensive cluster formation? The tails of NGC 3256 are very similar to
other tails that do not show such large numbers of clusters; all show
large numbers of inner region clusters; all are rich in atomic gas;
all have dynamical ages from 400-800 Myr (with NGC 3256 and NGC 3921
at the lower range of that scale).  However, NGC 3256 is extremely
bright in the X--ray and far--IR, and also has the largest number of
clusters in its inner regions.   

From the far infrared luminosities of each pair, we calculated the
global star formation rates using the relation of
\citet{Kennicutt1998}: $SFR=4.5\times10^{-44}L_{\rm
FIR}$~M$_{\odot}$~yr$^{-1}$.  This conversion assumes that the dust
re-emits all of the young starlight ($\tau >> 1$), and that the dust
heating is dominated by young stars, with ages of order $10^8$~years
or less.  As shown in Table~\ref{tab:sfr}, NGC~3256 stands out clearly
with a star formation rate several times that of the Antennae and NGC
7252, the next closest pairs.  Though these rates are calculated from
the integrated infrared luminosities rather than locally for the tidal
tails, the overall enhancement of star formation in NGC~3256 may be a
prerequisite for the efficient production of star clusters.  Also, the
star formation rate is related to the current burst of star formation
and might not indicate the conditions in the merger when the star
clusters in the tails formed, but perhaps the higher star formation
rate of NGC 3256 indicates a higher global molecular gas content that
would aid in more efficient star cluster formation.

We should consider whether the specific frequency of young clusters is
large for NGC 3256, or whether the number of young clusters is what
would be expected based upon the luminosity of the tail region.  From
\citet{Whit95} and \citet{goud01}, we can compare the specific
frequency of young clusters forming in the central regions of mergers
with $S_{young}$ of our tidal tails.  We calculated the $M_V$ of the
tidal tails by using the IRAF task IMSTAT at several typical locations
within the tail, and subtracted the background, measured at several
locations outside the tail.  The Western tail of NGC 3256 has a very
large specific frequency, $S_{young}=2.5$, comparable to
$S_{young}\footnote{This number corresponds to $S$ in
\citet{Whit95}.}\sim2$ found in the central regions of NGC 4038/39
\citep{Whit95} and $S_{young}\footnote{This number corresponds to
$S_N$ in \citet{goud01}.}=1.7$ found in NGC 1316 \citep{goud01}.  Clearly,
calculation of $S_{young}$ is complicated by destruction of clusters,
fading, etc.  However, it is still quite interesting that one of these
tails hosts clusters, that are as luminous, relative to stars as in
central regions of merging galaxies.

Another difference between NGC 3256 and the other pairs, is that,
unlike the other three systems, NGC 3256 does not have a prominent
tidal dwarf galaxy associated with the tip of its tidal tail either in
the optical image or in {\HI} maps.  A hypothesis, based on our
admittedly small sample of six tidal tails, is that tails with
prominent tidal dwarfs form fewer clusters than tails without such
dwarfs.  If we also consider the ACS early release observations of UGC
10214 and NGC 4676, they follow a similar trend.  Although no
correction was made for background contamination, UGC 10214 (which has
a tidal dwarf candidate) has fewer star clusters in its tail than NGC
4676 (which does not host a tidal dwarf candidate)
\citep{deGrijs03,Tran03}.  However, UGC 10214 is thought to be a
disturbed spiral with a single long tidal tail, a different
environment from the merging pairs of spiral galaxies that comprise
our sample and NGC 4676.  UGC 10214 also hosts many young star
clusters in the bright blue clump at the mid-point of its tail with
ages from $\sim3-10$~Myr \citep{Tran03}.

\citet{elm97} showed that large mass clumps and high specific kinetic
energies lead to a higher efficiency for star formation because of an
increased binding energy and resistance to disruption. Perhaps the
details of the interaction (i.e. the mass of the perturber, the
orbital properties of the two galaxies as they merge, the gas content
of the parent galaxy, the dark matter distribution, and the velocity
of the perturber) influence the process of star formation in the
debris.  We hypothesize that global characteristics of the encounters
affect the local conditions, conspiring so that star clusters form
along the tail or within a dwarf galaxy in the tail, but not both.

\acknowledgments
We thank the referee for a number of helpful comments and
suggestions. Support for this work was provided by grant STSI NASA
GO-07466.01-96A from the Space Telescope Science Institute, which is
operated by AURA, Inc., under NASA contract NAS5--26555.  Additional
support was provided by the National Science Foundation under grant
AST-0071223.  KAK was also supported by an NSF REU Supplement.

%%%%%%%%%%%%%%%%%%%%%%%%%%%%%%%%%%%%%%%%%%%%%%%%%%%%%%%%%%%%%%%%%%%%%%%%%%%%%%%%%%

%%%%%%%%%%%%%%%%%%%%%%%%%%%% figures %%%%%%%%%%%%%%%%%%%%%%%%%%%%%%%%%%%%%%%%%%

\onecolumn

\section*{Figures}

\placefigure{fig1}

\clearpage

\placefigure{fig2}

\clearpage

\placefigure{fig3}

\clearpage

\placefigure{fig4}

\clearpage

\placefigure{fig5}

\clearpage

\placefigure{fig6}

\clearpage

\placefigure{fig7}

\clearpage

\placefigure{fig8}

\clearpage

\placefigure{fig9}

\clearpage

\placefigure{fig10}

\clearpage

\placefigure{fig11}

\clearpage

\placefigure{fig12}

\clearpage

\placefigure{fig13}

\clearpage

\placefigure{fig14}

\clearpage

\placefigure{fig15}

\clearpage

%%%%%%%%%%%%%%%%%%%%%%%%%%%% tables %%%%%%%%%%%%%%%%%%%%%%%%%%%%%%%%%%%%%%%%%%%
\newpage

\input{Knierman.tab1.tex}
%\label{tab:obs}

\newpage
\input{Knierman.tab2.tex}
%\label{tab:prop}

\newpage
\input{Knierman.tab3.tex}
%\label{tab:summary}

\newpage
\input{Knierman.tab4.tex}
%\label{tab:sfr}

\end{document}

%% file: Knierman.tab1.tex
%\documentstyle[11pt,aaspp4]{article}
%\pagestyle{empty}
%\begin{document}

\begingroup
\begin{deluxetable}{lccc}
\footnotesize
\tablecaption{Journal of Observations}
%\vglue -0.5in
%\tablenum{1}
\tablewidth{6in}
\tablewidth{0pt}
\tablehead{\colhead{Tidal Tail} & \colhead{Filter} & \colhead{Time} &
\colhead{Date} \\
\colhead{} & \colhead{} & \colhead{s} & \colhead{}} 
\startdata
NGC$4038$ &  F555W & 2000 & 22 Feb 99 \\
NGC$4038$ &  F814W & 1800 & 22 Feb 99 \\
\hline 
NGC$3256$W &  F555W & 1000 & 24 Mar 99 \\
NGC$3256$W &  F814W & 1000 & 24 Mar 99 \\
\hline
NGC$3256$E &  F555W & 1000 & 11 Oct 99 \\
NGC$3256$E &  F814W & 1000 & 11 Oct 99 \\
\hline
NGC$3921$S &  F555W & 1800 & 30 Apr 99 \\
NGC$3921$S &  F814W & 1600 & 30 Apr 99 \\
\hline
NGC$7252$W & F555W & 2000 & 18 Nov 98 \\
NGC$7252$W & F814W & 1800 & 18 Nov 98 \\
\hline
NGC$7252$E & F555W & 2000 & 29 Aug 99 \\
NGC$7252$E & F814W & 1800 & 29 Aug 99 \\
\enddata
\label{tab:obs}
\end{deluxetable}
\endgroup

%\end{document}

%% file: Knierman.tab2.tex
\def\kms{\hbox{km~s$^{-1}$}}
\def\msun{${\rm M}_{\odot}$}
\def\HI{\hbox{{\rm H}\kern 0.1em{\sc i}}}
%\pagestyle{empty}
%\begin{document}

\begingroup
\begin{deluxetable}{lcccccccc}
\footnotesize
\tablecaption{Summary of Tidal Tail Properties}
%\tablenum{2}
\tablewidth{6in}
\tablewidth{0pt}
\tablehead{\colhead{Tidal Tail} & \colhead{Vel} & \colhead{Dist. Mod.} & \colhead{$A_B$\tablenotemark{a}} & \colhead{$l$} & \colhead{$\Delta l$} &
\colhead{$M_{V,50\%}\tablenotemark{b}$} & \colhead{$M_{HI,tail}$} &
\colhead{Tail Age}\\
\colhead{} & \colhead{\kms} & \colhead{mag} & \colhead{mag} &
\colhead{kpc} & \colhead{kpc} & \colhead{mag} & \colhead{\msun} &
\colhead{Myrs}}
%\hline
\startdata
NGC$4038$ & $1439$ & $31.4$ & $0.200$ & $30$ & $20$ & $-5.8$ & $2.0\times10^8$
& 420 \\
\hline 
NGC$3256$W & $2820$ & $32.9$ & $0.524$ & $15$ & $40$ & $-7.5$ & $2.2\times10^9$ &
400 \\
NGC$3256$E & $2820$ & $32.9$ & $0.524$ & $15$ & $30$ & $-7.5$ & $1.4\times10^9$ &
400 \\
\hline
NGC$3921$S & $6021$ & $34.5$ & $0.062$ & $0$ & $45$ & $-8.7$ & $4.3\times10^9$ &
460 \\
\hline
NGC$7252$W & $4828$ & $34.0$ & $0.130$ & $20$ & $50$ & $-8.1$ & $2.2\times10^9$ &
730 \\
NGC$7252$E & $4828$ & $34.0$ & $0.130$ & $0$ & $60$ & $-8.1$ & $1.5\times10^9$ &
730 \\
%\hline
\enddata

\tablenotetext{a}{From Schlegel {\etal} (1998).}
\tablenotetext{b}{Limiting magnitude for $\sim 50$\% completeness.}
\label{tab:prop}
\end{deluxetable}
\endgroup

%\end{document}

%% file: Knierman.tab3.tex
%\documentstyle[11pt,aaspp4]{article}
%\pagestyle{empty}
%\begin{document}

\begingroup
\begin{deluxetable}{lccccccccc}
\tabletypesize{\footnotesize}
\tablecaption{Source Densities In and Out of Debris}
%\vglue -0.5in
%\tablenum{3}
\tablewidth{6in}
\tablewidth{0pt}
\tablehead{\colhead{Tidal Tail} & \colhead{Pixel Size} & \colhead{$N_{out}$} & \colhead{$f_{out}$} & \colhead{$N_{out}/$Area} &
\colhead{$N_{in}$} & \colhead{$f_{in}$} & \colhead{$N_{in}/$Area} & \colhead{Surplus} &
\colhead{Spec. Freq.}\\
\colhead{} & \colhead{pc} & \colhead{}& \colhead{} &
\colhead{kpc$^{-2}$} & \colhead{} & \colhead{} & \colhead{kpc$^{-2}$}
& \colhead{} & \colhead{}}
%\hline
\startdata
NGC$4038$ & $8.92$ & $1$ & $0.501$ & $0.013\pm0.013$ & $1$ & $0.499$ &
$0.013\pm0.013$ & $0.000\pm0.018$ & $0.0$ \\
\hline 
NGC$3256$W & $18.2$ & $12$ & $0.677$ & $0.028\pm0.008$ & $28$ & $0.323$ &
$0.136\pm0.026$ & $0.108\pm0.027$ & $2.5$ \\
NGC$3256$E & $18.2$ & $4$ & $0.538$ & $0.012\pm0.006$ & $9$ & $0.462$ & 
$0.031\pm0.010$ & $0.019\pm0.012$ & $0.4$ \\
\hline
NGC$3921$S & $38.8$ & $7$ & $0.751$ & $0.003\pm0.001$ & $9$ & $0.249$ &
$0.013\pm0.004$ & $0.010\pm0.004$ & $0.2$\\
\hline
NGC$7252$W & $31.1$ & $5$ & $0.623$ & $0.004\pm0.002$ & $13$ & $0.377$ & 
$0.019\pm0.005$ & $0.015\pm0.005$ & $1.1$ \\
NGC$7252$E & $31.1$ & $13$ & $0.618$ & $0.011\pm0.003$ & $7$ & $0.382$ &
$0.010\pm0.004$ & $-0.001\pm0.005$ & $-0.2$ \\ 
%\hline
\enddata
\label{tab:summary}
\end{deluxetable}
\endgroup

%\end{document}

%% file: Knierman.tab4.tex
%\documentstyle[11pt,aaspp4]{article}
%\pagestyle{empty}
%\begin{document}

\begingroup
\begin{deluxetable}{lcc}
\tabletypesize{\footnotesize}
\tablecaption{Star Formation Rates of Merging Pairs}
%\vglue -0.5in
%\tablenum{4}
\tablewidth{6in}
\tablewidth{0pt}
\tablehead{\colhead{Pair} & \colhead{F$_{FIR}$} & \colhead{SFR} \\
\colhead{} & \colhead{erg cm$^{-2}$ s$^{-1}$} & \colhead{M$_\odot$ yr$^{-1}$}}
%\hline
\startdata
NGC$4038/9$\tablenotemark{a} & $2.6\times10^{-9}$ & $5.2$ \\
\hline 
NGC$3256$\tablenotemark{b} & $4.32\times10^{-9}$ & $33.4$ \\
\hline
NGC$3921$\tablenotemark{b} & $(2.69-5.25)\times10^{-11}$ & $1.9$ \\
\hline
NGC$7252$\tablenotemark{c} & $2.37\times10^{-10}$ & $5.4$ \\
 
\enddata

\tablenotetext{a}{From Ranalli, Comastri, \& Setti (2003).}
\tablenotetext{b}{From Moshir et al. (1990).}
\tablenotetext{c}{From Knapp et al. (1989).}

\label{tab:sfr}
\end{deluxetable}
\endgroup

%\end{document}